# The helium-core mass at the helium flash in low-mass red giant stars: observations and theory

M. Catelan, J. A. de Freitas Pacheco[1] and J. E. Horvath

Instituto Astronômico e Geofísico, Universidade de São Paulo

## ABSTRACT

The method developed by Raffelt (1990a,b,c) to estimate a possible increase in the standard values of the helium-core mass at the tip of the red giant branch, $M_c$, from properties of the color-magnitude diagrams of Galactic globular clusters is employed. In the present study, we revise and update Raffelt's database, including also constraints from RR Lyrae pulsation, and find that a small increase, by $\Delta M_c \approx 0.01 \pm 0.015\,M_\odot$, cannot be ruled out with the present data and evolutionary models. Our new upper limits on $\Delta M_c$ are less restrictive than those previously obtained by Raffelt, as are the corresponding constraints on novel astroparticle phenomena which may affect the evolution of low-mass red giants. Within the estimated uncertainties, however, the standard values of $M_c$ may also be acceptable. Raffelt's method does not rule out a low envelope helium abundance in globular cluster giants, though again the standard values are compatible with the available constraints. The influence of a non-solar ratio for the $\alpha$-capture elements upon these results is also investigated. In addition, we review several aspects of the input physics employed in red giant stellar evolutionary calculations, with the purpose of evaluating possible sources of uncertainty in the value of the helium-core mass at the helium flash that is obtained from evolutionary computations, such as: heat conduction by electrons in the degenerate core; Coulomb effects upon the Equation of State; triple-$\alpha$ reaction rates and screening factors; neutrino emission rates, both standard and enhanced by a possible non-zero magnetic moment; stellar rotation; microscopic element diffusion; and energy losses by axions and Weakly Interacting Massive Particles.

---

[1] Observatoire de la Côte d'Azur, B.P. 229, Nice Cedex 04, 06304 France.

## 1. Introduction

The determination of the mass of the degenerate helium core of low-mass red giant stars at the tip of the red giant branch (RGB), $M_c$, is a very important astrophysical problem for a variety of reasons. $M_c$ controls the luminosity of the star near the RGB tip, which is important for the determination of the rate of mass loss and thus the horizontal-branch (HB) morphology (e.g., Fusi Pecci & Renzini 1976). The luminosity of the RGB tip is important in distance-scale arguments as well (Lee, Freedman, & Madore 1993). Also, $M_c$ controls all the subsequent evolution on the HB and beyond, and is a crucial factor determining the sense of evolution of the RR Lyrae variables (e.g., Sweigart & Gross 1976): increased $M_c$ values would drive redward evolutionary paths for RR Lyrae variables in Oosterhoff type II globular clusters (GCs), and thus help account for the Sandage period-shift effect (Sandage 1982, 1990, 1993) and related problems in terms of the evolutionary scheme of Lee, Demarque, & Zinn (1990), as recently emphasized by Catelan (1992a,b, 1993a,b). Perhaps even more importantly, the luminosity of the HB depends markedly on $M_c$, and so does the distance (and hence age) scale that is most commonly employed for the GCs in the halo of the Galaxy and nearby systems.

The evolutionary $M_c$ values obtained by Sweigart & Gross (1978, hereafter SG78) have been confirmed by several authors (e.g., Straniero & Chieffi 1991; VandenBerg 1992a,b). However, some computations have resulted in values which are not in clear agreement with the SG78 results (e.g., Harpaz & Kovetz 1988; Alongi et al. 1991; Raffelt & Dearborn 1987). In particular, Mazzitelli (1989, hereinafter M89) has claimed the SG78 values to be systematically too small by $\approx 0.02\,M_\odot$, which he attributed to the inadequacy of the "shell-shifting" technique (Härm & Schwarzschild 1966) for shifting the hydrogen-burning shell and advancing the chemical composition profile in an evolving RGB star and to an inappropriate choice of input physics by SG78. Very recently, Sweigart (1994) has analyzed in great detail the numerical aspects of the problem, and found that the $M_c$ values do not hinge strongly on the shell-shifting algorithm employed. On the other hand, Sweigart suggests that the M89 prescriptions for implicitly advancing the chemical composition profile may lead to an *overestimate* of the helium-core mass, and thus should not be employed.

Even though the numerical aspects of the problem of RGB computations have been put on a solid basis by Sweigart's (1994) analysis, not so much attention has been devoted to the input physics. In particular, Sweigart has not addressed the non-standard effects which may play a rôle in determining the core mass at the helium flash.

It is the purpose of the present report to give an updated account of several aspects of the input physics that controls the core mass at the helium flash. We shall divide the present study into four main sections. In Sect. 2, Raffelt's (1990a,b,c) method to evaluate a possible increase in $M_c$ is critically readdressed, on the basis of recent observational and theoretical results. The method suggests that a $\Delta M_c > 0$ cannot be ruled out with the present data, and may actually be favored by them. In Sect. 3, relevant standard ingredients of RGB evolutionary calculations are addressed: electron heat conductivities, the Equation of State (EOS), reaction rates, and standard neutrino losses. In Sect. 4, the present status of important ingredients of non-standard evolutionary calculations, stellar rotation and microscopic diffusion, is critically discussed. In Sect. 5, exotic stellar energy loss mechanisms are considered, as caused by the presence of a non-zero neutrino magnetic moment, a "light" axion, and Weakly Interacting Massive Particles (WIMPs). Finally, our main conclusions are summarized in Sect. 6.

## 2. $\Delta M_c$ from the observations and stellar evolution: Raffelt's method revisited

As already stated, the value of $M_c$ has a marked impact on the luminosities of the RGB tip stars and the RR Lyrae variables. Also, $M_c$ controls the lifetimes of the stars on the RGB and HB evolutionary phases and the RR Lyrae pulsation properties.

Recently, Raffelt (1990a,b,c) has employed the available observational data and stellar evolution models to evaluate $\Delta M_c \equiv M_c^{\mathrm{actual}} - M_c^{\mathrm{standard}}$. After an extensive analysis, Raffelt (1990a) arrived at two expressions for the increase in $M_c$ values [his eqs. (37) and (38)], and from these concluded that the increase in $M_c$ cannot be very large, with a likely $1 - \sigma$ limit of $\Delta M_c \approx 0.02\,M_\odot$ [cf. his eq. (39)].

A revision of Raffelt's (1990a,b,c) analysis at the present time is appropriate for several reasons. First, the author has employed for the RR Lyrae variables a mean magnitude of $\langle M_V \rangle = 0.68$ mag, which several new lines of evidence now suggest may be too faint (e.g., Saha et al. 1992; Catelan 1992a,b, 1993a,b; Simon & Clement 1993; Cacciari & Bruzzi 1993; Sandage 1993 and references therein; Fernley 1994; see, however, Storm, Carney, & Jones 1994). In particular, the distance modulus of the Large Magellanic Cloud (LMC) determined from independent distance indicators, such as the analysis of the SN1987A circumstellar ring (e.g., Panagia et al. 1991; Crotts, Kunkel, & Heathcote 1995), gives strong support to this result (Walker 1992a). Some of these points have been recently addressed by Fernley (1994) and van den Bergh (1995). Second, RR Lyrae pulsation has not been addressed by Raffelt, even though it is known that the core mass influences the RR Lyrae properties in a very significant way (see, e.g., Catelan 1992a,b, 1993a,b and Castellani & De Santis 1994 for recent discussions). Finally, in Raffelt's original analysis several theoretical relations which are strictly applicable only to zero-age HB stars have been employed, while his sample of GCs includes several objects with extreme HB morphologies, where these variables may be well-evolved (Lee 1990; Catelan 1992a,b, 1993a,b; Caputo et al. 1993).

We thus update Raffelt's (1990a,b,c) analysis in the present section, taking these problems into consideration.

## 2.1. Theoretical relations

In Raffelt's (1990a,b,c) study, extensive use has been made of the relations derived by Buzzoni et al. (1983), which aimed at a semi-empirical determination of the helium abundance in GC stars through the so-called "$R$-method" (Iben 1968b). We shall similarly employ those relations in the present analysis.

We assume that the helium flash does not induce mixing of metals and helium to the star's envelope (Iben & Renzini 1984; Renzini & Fusi Pecci 1988). This assumption has yet to be checked in detail by self-consistent evolutionary and hydrodynamic computations, which however are strongly dependent on initial and boundary conditions, as well as on the treatment of convective energy transport in the core (see, e.g., Deupree & Wallace 1987 for a discussion).

### 2.1.1. The RR Lyrae magnitude level

The dependence of the HB absolute visual magnitude at the RR Lyrae level on the envelope helium abundance $Y$, metallicity $Z$, and $\Delta M_{\rm c}$, is given by the following relation:

$$Y = 0.561 - 0.285 \langle M_{\rm bol}^{\rm RR} \rangle + 0.046 \log Z - 0.285 \, \Delta_{\rm RR} - 2.083 \, \Delta M_{\rm c}, \tag{1}$$

where $\Delta_{\rm RR}$ represents the deviation of the RR Lyrae luminosities from the zero-age predictions at $\log T_{\rm eff} = 3.85$. A bolometric magnitude $M_{\rm bol} = 4.74$ mag has been adopted for the sun. This follows from the analysis of recent satellite data, as summarized by Guenther et al. (1992), Sackman, Boothroyd, & Kraemer (1993), and Bahcall & Pinsonneault (1995), and the formula of Lang (1986) to convert luminosities into bolometric magnitudes. In Raffelt's (1990a) original expression [his eq. (9)], $M_{\rm bol}^{\odot} = 4.72$ mag had been used instead.

### 2.1.2. The brightness of RGB tip stars

In order to estimate the *absolute* magnitudes of stars at the tip of the RGB, an assumption is required as to the distance moduli of the corresponding clusters (Frogel, Persson, & Cohen 1983).

For our present purposes, we determine the magnitude *difference* between the HB at the RR Lyrae level and the RGB tip, $\Delta M_V^{\rm tip-RR}$, and compare the observed values with the theoretical predictions (Raffelt 1990a). For the latter, following Buzzoni et al. (1983) and Raffelt (1990a), we find

$$Y = 1.455 - 0.227 \, \Delta M_V^{\rm tip-RR} + 0.089 \log Z - 0.227 \, \Delta_{\rm RR} + 0.909 \, \Delta M_{\rm c}. \tag{2}$$

### 2.1.3. Star number counts and the R-method

From the theoretical relations provided by Buzzoni et al. (1983) for the lifetimes of giants and HB stars, the envelope helium abundance may be estimated as follows:

$$Y = 0.146 + 0.437 \log R - 0.013 \log Z - 0.144 \, \Delta_{\mathrm{RR}} + 0.306 \, \Delta M_{\mathrm{c}}, \qquad (3)$$

where $R$ represents the ratio between the number of RGB stars brighter than the HB level and the total number of HB stars, including the RR Lyrae variables.

### 2.1.4. Pulsation properties of RR Lyrae stars

The value of $M_c$ controls the luminosity, mass, and temperature of a star inside the instability strip (e.g., Sweigart & Gross 1976), and thus, through the period-mean density relation, also its pulsation period, with an increase in $M_c$ generally leading to an increase in the period.

Comparison between the standard models and the observations shows that the computed periods appear too small. Catelan (1992a,b, 1993a,b) has suggested that a systematic error in the evolutionary $M_c$ values may be responsible for this. However, the dependence of the periods on the effective temperature is large, which complicates the comparison between the models and the observations. On the observational front, small errors in the reddening can be of importance in "fixed-temperature" arguments (cf. Caputo & De Santis 1992 and references therein). On the theoretical side, the location of the blue and (especially) red edges of the instability strip is another delicate issue, and great theoretical effort is now being devoted toward their accurate evaluation (Bono & Stellingwerf 1994 and references therein). Since the RR Lyrae "mass-to-light ratio," $A \equiv \log L - 0.81 \log M$, may be computed from the observational data in a reddening-independent way (Caputo & De Santis 1992), it may be more safely compared with the theoretical predictions at a fixed effective temperature for stars close to the zero-age HB.

We have obtained the theoretical predictions for the zero-age HB mass-to-light ratio $A$ at an effective temperature $\log T_{\mathrm{eff}} = 3.85$ from the models of Sweigart, Renzini, & Tornambè (1987). The results displayed in their Table 4 may be fitted by the following equations:

$$\log M = -1.122 - 0.082 \log Z + 0.191 \, Y_{\mathrm{MS}} + 1.294 \, M_{\mathrm{c}}, \qquad (4)$$

and

$$\log L = -0.451 - 0.026 \log Z + 1.96 \, Y_{\mathrm{MS}} + 3.197 \, M_{\mathrm{c}}, \qquad (5)$$

where $M$ and $L$ are the stellar mass and luminosity, respectively, both in solar units. $Y_{\mathrm{MS}}$

represents the helium abundance prior to the main-sequence phase. With the aid of equations (1) and (6) from Raffelt 1990a, we finally obtain

$$Y = -1.146 + 0.750\,A_{3.85} - 0.010\,\log Z - 1.548\,\Delta M_c. \tag{6}$$

## 2.2. Observational relations

### 2.2.1. The RR Lyrae magnitude level

For the absolute magnitude of the HB at the (mean) RR Lyrae level, we employ Walker's (1992a) recent analysis of old GCs in the LMC with significant numbers of RR Lyrae variables. Assuming $(m - M)_0 = 18.50$ for the LMC (e.g., Panagia et al. 1991), Walker (1992a) finds

$$\langle M_V^{\mathrm{RR}} \rangle = 0.15\,[\mathrm{Fe/H}] + 0.73. \tag{7}$$

In the present work, we adopt the revised distance modulus for the LMC recently obtained by Crotts et al. (1995) through detailed analysis of the geometry of the SN1987A circumstellar envelope, $(m - M)_0 = 18.58 \pm 0.13$, and convert the $\langle M_V^{\mathrm{RR}} \rangle - [\mathrm{Fe/H}]$ relation to the corresponding $\langle M_{\mathrm{bol}}^{\mathrm{RR}} \rangle - [\mathrm{Fe/H}]$ one with the aid of Dorman's (1992) prescriptions (cf. Sandage 1993), so that

$$\langle M_{\mathrm{bol}}^{\mathrm{RR}} \rangle = 0.184\,[\mathrm{Fe/H}] + 0.68. \tag{8}$$

We adopt an uncertainty in values obtained from this equation of $\pm 0.16$ mag.

It is important to note that, according to the data compiled in Walker's (1992b) Table 7, the Walker 1992a cluster sample is *not* biased in an evolutionary sense, since it does not appear to contain objects with extreme HB morphologies. Also, the adopted distance modulus for the LMC appears to be supported by several independent distance indicators, such as Cepheids and Miras, as discussed by Walker (1992a).

### 2.2.2. The brightness of RGB tip stars

When studying the brightness difference between the RGB tip and the HB at the RR Lyrae level, Raffelt (1990a), in his detailed analysis, has included some objects which have too few RR Lyrae variables and thus cannot be safely employed in a comparison with the zero-age HB relations within the instability strip. This is due to the possibility that, in clusters with too few RR Lyrae variables, these stars are evolved *from the blue (or red) zero-age HB*, and thus not representative of the zero-age HB at the RR Lyrae level (Lee et al. 1990; Lee 1990; Catelan 1992a,b, 1993a,b;

Caputo et al. 1993). Thus, we have repeated Raffelt's analysis excluding the clusters in his Table 2 with HB morphologies outside the range $-0.7 \leq (B-R)/(B+V+R) \leq +0.7$, where $B$, $R$, and $V$ are the number of stars to the blue, to the red, and within the instability strip, respectively (data taken from Lee, Demarque, & Zinn 1994). NGC 2808, which has the peculiarity of a dramatic *internal* second-parameter effect, has also been removed. We are left with the following objects: NGC 1261, NGC 1851, NGC 3201, NGC 5272, NGC 5904, NGC 6121, and NGC 7006. The metallicities have also been revised, following Lee et al. (1994). According to Raffelt's equations (22) to (25), and adopting $\sigma_{\rm obs} = 0.1$ mag, we then find

$$\langle \Delta M_V^{\rm tip-RR} \rangle = 4.09 \pm 0.15, \tag{9}$$

for a mean metallicity $\langle [{\rm Fe/H}] \rangle = -1.47 \pm 0.15$.

### 2.2.3. Star number counts and the R-method

Under the same restrictions as in the previous subsection, the data in Table 5 of Raffelt (1990a) associated to the number ratio counts have been reanalyzed. Clusters satisfying the HB morphology criterion were NGC 1851, NGC 3201, NGC 4147, NGC 5272, NGC 5904, and NGC 6121. We find that, for such clusters,

$$\langle \log R \rangle = 0.123 \pm 0.037, \tag{10}$$

for a mean metallicity $\langle [{\rm Fe/H}] \rangle = -1.49 \pm 0.17$.

### 2.2.4. Pulsation properties of RR Lyrae stars

Employing a relation analogous to Caputo & De Santis's (1992) equation (12) (cf. Catelan 1995) in connection with the observational data for M3 ([Fe/H] $= -1.66$) provided by Sandage (1990) in his Table 5, we find that

$$A_{3.85} = 1.80 \pm 0.01. \tag{11}$$

In obtaining this value, we have employed Catelan's (1993a) procedure to search for the lower envelope of a distribution. In the present case, we assume that the zero-age HB corresponds to the lower envelope thus obtained. Sandage (1990) has argued that for this cluster this is probably a good approximation, as supported by synthetic HB models (cf. Catelan 1993a and references therein). Independent application of this method to the cluster NGC 3201 confirms this result. We caution, however, that the above error bar does not include systematic uncertainties, as might be present if the temperature scale employed by Caputo & De Santis (1992) is not fully adequate. Further details will be given elsewhere (Catelan 1995).

## 2.3. Theory and observations confronted

To compare the theoretical and observational relations obtained above, it is necessary to evaluate the quantity $\Delta_{\mathrm{RR}}$ appearing in some of them. Raffelt (1990a,b,c) has neglected such terms in his analysis. In the present work, we follow the discussion by Catelan (1992a), where the variation of the HB magnitude dispersion at the RR Lyrae level as a function of [Fe/H] in Sandage's (1990) sample of GCs has been studied.

Since

$$\Delta M_V^{\mathrm{RR}} = 0.751 + 0.235 \, [\mathrm{Fe/H}] \tag{12}$$

is Catelan's (1992a) prescription for the run of the HB thickness with cluster metallicity, and since

$$\langle \eta \rangle = (41 \pm 15)\% \tag{13}$$

is the mean "concentration parameter" (measuring the mean deviation of the variables from the lower envelope of the distribution) for those clusters studied by him with intermediate HB morphology, it follows that

$$\Delta_{\mathrm{RR}} \simeq 0.31 + 0.10 \, [\mathrm{Fe/H}]. \tag{14}$$

We now insert these results into the theoretical relations given in Sect. 2.1. For the RR Lyrae magnitudes, for consistency with the observational data for $R$ and $\Delta M_V^{\mathrm{tip-RR}}$, we adopt the value given in equation (8) for a metallicity [Fe/H] $= -1.48 \pm 0.16$, and find

$$Y = (0.250 \pm 0.049) - 2.083 \, \Delta M_c, \qquad \text{(for } M_V^{\mathrm{RR}}) \tag{15}$$

where $\sigma(M_{\mathrm{bol}}^{\mathrm{RR}}) = 0.16$ mag and $\sigma(\Delta_{\mathrm{RR}}) = 0.06$ mag have been assumed. In a similar vein, from the number ratio counts ($R$-method), we find

$$Y = (0.218 \pm 0.018) + 0.306 \, \Delta M_c. \qquad \text{(for } R) \tag{16}$$

From the magnitude difference between the HB and the RGB tip, one finds

$$Y = (0.207 \pm 0.039) + 0.909\,\Delta M_{\mathrm{c}}. \qquad (\text{for } \Delta M_V^{\mathrm{tip-RR}}) \tag{17}$$

Finally, the mass-to-light parameter $A$ yields

$$Y = (0.238 \pm 0.008) - 1.548\,\Delta M_{\mathrm{c}}. \qquad (\text{for } A_{3.85}) \tag{18}$$

Each of these four equations defines a band in the $Y - \Delta M_{\mathrm{c}}$ plane, as plotted in Fig. 1. As can be seen, all four bands intersect at a relatively narrow "compatibility region" in the middle of the diagram, around $(Y, \Delta M_{\mathrm{c}}/M_\odot) \simeq (0.221 \pm 0.017, +0.011 \pm 0.014)$, which defines the preferred solution for the envelope helium abundance and the increase in $M_{\mathrm{c}}$. Regions depicted in lighter gray tones in this diagram define progressively less likely solutions.

Analysis of this diagram reveals that, even though the data are generally compatible with the standard $M_{\mathrm{c}}$ values, an increase by $\approx 0.010 - 0.015\,M_\odot$ (or possibly larger) cannot be ruled out, and in fact may appear to be favored. It is interesting to note that the narrow band running from upper left to lower right in this diagram, coming from equation (18), is of great importance in constraining the allowed solutions. Such a constraint, obtained from the analysis of RR Lyrae pulsation properties, had not been previously taken into consideration in the ignition argument (Raffelt 1990a,b,c). Systematic uncertainties were not included in the analysis of $A_{3.85}$, though, so that the actual "compatibility region" may be slightly wider. In any event, our $1 - \sigma$ limit for $\Delta M_{\mathrm{c}}$ is appreciably larger than Raffelt's, posing less stringent constraints on non-standard physical phenomena that may be affecting the evolution of giant stars (cf. Sect. 5).

Lower envelope helium abundances are also consistent with Fig. 1. As already mentioned, a low $Y$ is favored by the evolutionary interpretation of the Sandage period-shift effect. A more detailed analysis of the helium abundance problem can be found in Catelan 1992b, where the difficulties in accounting for an abnormally low $Y$ in GCs were discussed. In particular, a low helium abundance may not appear compatible with primordial $^4$He nucleosynthesis arguments, both theoretical and observational (e.g., Pagel et al. 1992; Pagel & Kazlauskas 1992; Olive, Steigman, & Walker 1991; Olive & Steigman 1994). Fig. 1 may also suggest that a slightly larger $Y$ would imply a $\Delta M_{\mathrm{c}}$ closer to zero. Furthermore, relaxation of the pulsation bounds would allow considerably higher $Y$ values. At any rate, it must be emphasized that a low *envelope* helium abundance in an evolved star does not necessarily imply a low *initial* abundance for this star, since diffusive mechanisms may decrease the envelope abundance along a star's lifetime.

## 2.4. Influence of an overabundance of the $\alpha$-elements

The above estimates have been made assuming solar-scaled abundances for the GC stars, i.e., $[\alpha/\mathrm{Fe}] = 0$. However, it is well known (see Wheeler, Sneden, & Truran 1989 for a review) that halo stars present an overabundance of the $\alpha$-elements, typically by a factor $\simeq 3$ (i.e., $[\alpha/\text{Fe}]$ $\approx 0.48$). Therefore, the above analysis may require modifications due to this specific effect.

Until now, the only extensive study dealing with the evolutionary implications of an overabundance of the $\alpha$-elements in low-mass stars is the one by Chieffi, Straniero, & Salaris (1991) and Salaris, Chieffi, & Straniero (1993). According to Chieffi et al., the $\alpha$-enhanced models may be obtained from the solar-scaled models on the basis of a straightforward rescaling of the latter in terms of the overall metal abundance $Z$, following a relation of the type

$$\log Z = [\text{Fe/H}] - 1.7 + \log(0.579\,f + 0.421), \tag{19}$$

where $f = 10^{[\alpha/\text{Fe}]}$. For the evolutionary stages that precede the HB phase, these results have been confirmed by Chaboyer, Sarajedini, & Demarque (1992).

Following the Chieffi et al. (1991) prescriptions, we have repeated the above analysis assuming $[\alpha/\text{Fe}] = 0.48$. Our results are summarized in Fig. 2. Even though the impact of the $\alpha$-enhancement upon the four individual relations between $Y$ and $\Delta M_c$ is quite significant — larger, in fact, than suggested by the Raffelt & Weiss (1992) analysis — the overall picture does not appear to change appreciably. In particular, the "compatibility region" now appears slightly narrower, being found around $(Y, \Delta M_c/M_\odot) \simeq (0.223 \pm 0.012, +0.008 \pm 0.011)$.

Although the agreement between the model predictions and all the available observational constraints does seem remarkably satisfactory, we emphasize that an increase in $M_c$ reaching perhaps as much as $0.015\,M_\odot$ cannot be ruled out. Since Sweigart (1994) has very convincingly argued against an increase in $M_c$ due to numerical problems in state-of-the-art evolutionary computations, it should be worthwhile to check the input physics in more detail.

## 3. Standard evolutionary effects, standard physics

In the present section, we shall analyze important ingredients of standard evolutionary calculations, aiming at an evaluation of their present uncertainties and the corresponding impact upon $\Delta M_c$.

### 3.1. Electron thermal conductivities in the interior of RGB stars

As one advances inward in the interior of an RGB star, electron conduction is progressively more efficient as an energy transport mechanism in comparison with radiative transfer, and becomes the dominant mechanism in the electron-degenerate helium core (see, e.g., the discussion accompanying Fig. 1 in Iben 1968a).

It is common practice, in stellar evolution calculations, to employ either the conductivities that have been presented by Hubbard & Lampe (1969, hereinafter HL69) — as in SG78 — or the more recent results by Itoh et al. (1983, hereinafter I$^3$M83) — as in M89. To our knowledge, however, a discussion of the hypotheses and uncertainties involved in the adoption of either formalism has not been presented in the literature. We suspect that this is due to the fact that, since the work by Iben (1968a), an understanding has grown that "electron conduction opacities in degenerate cores are so small anyway that the core is isothermal in any case" (quotation from Raffelt & Weiss 1992). However, Sweigart et al. (1987) have shown that, by multiplying conductive opacities by a factor $F_c = 0.5$, $M_c$ may increase by $0.0087 - 0.01\,M_\odot$ (cf. their Table 6). Since this is in the ballpark of the "compatibility regions" in Figs. 1 and 2, some exploration of this topic should be of interest.

"Conductive opacity," $\kappa_c$, is usually defined, in terms of the more familiar conductivity $\lambda_c$, as follows:

$$\kappa_c = \frac{4acT^3}{3\rho\lambda_c},\tag{20}$$

where a is the radiation density constant, c is the speed of light, $\rho$ is the density, and $T$ is the temperature. The usefulness of such a definition is that it places the heat conduction and radiative transfer equations in identical forms (e.g., Clayton 1983).

In Fig. 3, a comparison is made of the conductive opacities available in the literature for the physical conditions prevailing in the center of RGB stars. For illustrative purposes, we plot the conductive opacities along the RGB evolution of the $M = 0.9\,M_\odot$, $Y_{\rm MS} = 0.20$, $Z = 10^{-4}$ model of SG78, based upon the log $\rho_c$ and log $T_c$ values displayed in their Table 2. For convenience, we have assumed that $Y_c = 1$.

In Fig. 3a, the conductive opacities are directly displayed. The classical analysis of HL69 is represented both by interpolative equation (A4) of Iben 1975 (hereafter Ib75) and by direct linear interpolation on their published tables. Equation (A28) of Ib75 represents Canuto's (1970) relativistic results. An interpolative scheme has been suggested by Ib75 [his eqs. (A31) and (A32)], whereby the HL69 results are valid for log $\rho < 6$, and Canuto's for log $\rho > 6.3$. For the Yakovlev & Urpin (1980) case, we have adopted equation (19) of Blinnikov & Dunina-Barkovskaya (1992). Equation (7) of I$^3$M83 has been employed as representative of their analysis, and the quantum corrections suggested by Mitake, Ichimaru, & Itoh (1984) were evaluated on the basis of their equation (8). Finally, the contribution to conductive opacity due to electron-electron interactions, as calculated by Urpin & Yakovlev (1980), was obtained from equation (27) of Blinnikov & Dunina-Barkovskaya (1992), with the aid of equation (10) by Timmes (1992).

In Fig. 3b, the ratio of conductive opacities from the different authors to the Ib75 approximation to the HL69 results is shown. Several interesting features are apparent.

First, the Canuto (1970) formulation coincides closely with the Ib75 approximation to the HL69 results for $\rho \simeq 10^6 \, \mathrm{g \, cm^{-3}}$, so that the relativistic corrections proposed by this author are of no practical importance for the (essentially classical) core. As a consequence, the curve labeled "Hubbard & Lampe" matches almost perfectly the one obtained from the Ib75 proposed interpolation between Canuto's (1970) and HL69's results all across the diagram of Fig. 3a, which is the reason why the latter is not displayed. Similarly, the quantum corrections that have been obtained by Mitake et al. (1984) are also very small. Third, the Yakovlev & Urpin (1980) opacities are somewhat larger than those due to HL69. Last but not least, it is also noted that the electron-electron contribution to the conductive opacity, as prescribed by Urpin & Yakovlev (1980), may be quite significant near the RGB tip.

Conductive opacities, as displayed in Fig. 3, never disagree by more than $\simeq 45\%$. Caution is needed, however, before concluding that present uncertainties in $M_c$ of order $0.01 \, M_\odot$ from this source are an upper limit. In particular, it should be noted that:

- The formulation of I³M83 and Mitake et al. (1984) has been developed aiming at applications to the more extreme conditions characterizing white dwarfs and neutron stars. *It is not expected to be especially accurate for the interior of an RGB star.* In Fig. 3b, a curve has been drawn to show the values achieved by the parameter

$$\Gamma \equiv \frac{Z^2 e^2}{r k_B T}, \tag{21}$$

where e is the electron charge, $k_B$ is the Boltzmann constant, $Z$ is the atomic number, and $r$ is the ion-sphere radius, $r = (3/4\pi n_i)^{1/3}$, with $n_i$ the ion number density. $\Gamma$ measures the strength of the electrostatic interaction among the plasma ions. [A related quantity, $\Lambda$, is also displayed, and will be discussed in the next section.] For $\Gamma \gtrsim 178$, the ions form a period structure (i.e., a crystal), whereas for $\Gamma \lesssim 2$ a Boltzmann gas is still present. We see that $\Gamma$ never exceeds 0.81 in the RGB core. *The calculations of I³M83 and Mitake et al. (1984) are valid only for $2 \leq \Gamma \leq 160$, i.e., for the liquid phase,* as explicitly stated by these authors. Thus, one should keep in mind that employing their results for low-mass RGB stars actually involves an extrapolation to the Boltzmann gas conditions that better characterize the ions in such stars;

- To shed light on the uncertainties involved in such extrapolations, Canuto's (1970) relativistic results were extrapolated towards $\log \rho < 6$, as shown in Fig. 3. The agreement with the HL69 opacities clearly deteriorates for the lower densities;

- The Itoh et al. (1984) results should *not* be employed in RGB models, since they refer to even more extreme conditions, $171 \leq \Gamma \leq 5000$ (there is also a recent erratum to this paper which should be referred to). The recent analyses by Itoh & Kohyama (1993) and Itoh, Hayashi, & Kohyama (1993) again refer exclusively to the crystalline lattice phase, and the present authors are unaware of an extension of this work to the regions of interest for RGB calculations (see also Itoh 1992);

- In their compilation of input physics for white dwarf models, Blinnikov & Dunina-Barkovskaya (1992) have pointed out that the calculations of $I^3M83$ and Yakovlev & Urpin (1980) do not include the effects of electron-electron interactions on the conductivities. As already noted, the $e - e$ contribution to the opacity for the relevant conditions may be quite significant (cf. Fig. 3). This point may have been neglected in some recent evolutionary calculations. The $e - e$ and $e - i$ contributions to the opacity add linearly, in a first approximation;

- Nandkumar & Pethick (1984) have also carried out an analysis similar to the one by $I^3M83$, giving results for $\Gamma > 4$. They find that inclusion of physical ingredients neglected by the latter authors, related to the influence of the electrons upon the effective interactions among the ions and between ions and electrons, can lead to differences of a few percent in the conductivities at low densities. Inspection of their Fig. 1, however, discloses that the effect referred to is negligible at $\rho \sim 10^6 \, \mathrm{g \, cm^{-3}}$;

- The Ib75 approximation to the HL69 conductive opacities appears to underestimate the values obtained by direct interpolation on HL69's Table 2 in the regions of interest for RGB interior computations. Approximation formulas of this kind are sometimes employed in stellar evolution codes. Our comparison shows that, whenever possible, accurate interpolation routines should be employed to obtain the opacities from the original references.

From these considerations, a few additional remarks are needed. One of the major differences between the M89 and SG78 analyses is that in the former the more recent ($I^3M83$) conductive opacities were adopted. *If* the extrapolation carried out is appropriate, then Fig. 3 shows the conductive opacities of $I^3M83$ to be somewhat smaller than those obtained by direct interpolation on HL69's tables (but *larger* than the values obtained from Ib75's proposed approximation!). This being the case, the core mass values attained with the $I^3M83$ opacities should be slightly *larger* (i.e., by a few times $0.001 \, M_\odot$), since it would be easier for the helium core to cool and the helium flash would take place later. Part of this difference might be removed if the $e - e$ interaction were properly taken into account. Note, however, that the Urpin & Yakovlev (1980) $e - e$ curves in Fig. 3 are themselves extrapolations, since these authors have provided results only for $\rho > 10^6 \, \mathrm{g \, cm^{-3}}$. One run of Mazzitelli's models with the HL69 conductivities has indeed resulted in a smaller $M_c$, although it is difficult to isolate the impact of $\kappa_c$ upon this result since several other ingredients were also changed in the same run. Thus, the disagreement between M89 and SG78 may be partially ascribed to the particular formulations for conductivities employed by these authors.

It appears clear to us that useful work could still be developed in the field of conductive opacities with the purpose of estimating more precisely the value of $M_c$. In particular, evaluation of the inaccuracies involved in extrapolating the $I^3M83$ calculations toward $\Gamma < 1$ is needed. While this is not done, we are inclined to suggest that preference should be given to the HL69 results in population II evolutionary calculations.

To close this section, we note, after the discussion by Iben (1968a), that inaccuracies in *radiative* opacities near the core boundary may also influence $M_c$ values in an important way. This should also be kept in mind when updating the input physics in an evolution code.

### 3.2. Coulomb effects upon the EOS

Coulomb effects upon the EOS were not included in the SG78 computations. According to Harpaz & Kovetz (1988), Coulomb interactions lead to a reduction in $M_c$ by $\approx 0.005\, M_\odot$, as a consequence of a reduction in the pressure leading to higher core temperatures, and thus to a flash that takes place earlier. The effect is not very large because at the conditions prevailing in RGB cores the EOS is dominated by electron degeneracy.

Coulomb corrections, when implemented in stellar evolution codes, are generally based on the Debye & Hückel (1923) formalism or some modification thereof (see, for instance, Straniero 1988 for a discussion and references). However, according to Rogers (1994), the Debye-Hückel correction to the EOS is only valid for

$$\Lambda \equiv \frac{\sqrt{4\pi n}Z^3 \mathrm{e}^3}{(\mathrm{k_B}T)^{3/2}} \lesssim 0.2. \tag{22}$$

Outside this range, more detailed treatments are required. As shown in Fig. 3b, $\Lambda > 1.1$ in an RGB core, so that the Debye-Hückel formalism breaks down and more sophisticated analyses are warranted to accurately model the evolution of RGB properties.

Only very recently (Chaboyer & Kim 1995) has the detailed OPAL EOS (cf. Rogers 1994 and references therein) been incorporated into evolutionary computations. However, these authors could not evolve their models up the RGB due to numerical difficulties. Extending their analysis to the RGB phase should be of great importance for the determination of the actual influence of Coulomb effects upon the $M_c$ values attained at the flash. [2]

---

[2]In this regard, it is important to note that the determination of fully self-consistent $\Delta V$ ages with the OPAL EOS must await the computation of detailed RGB sequences, since HB evolutionary characteristics — luminosities in particular — depend on $M_c$, as discussed in Sect. 2. As an example, a decrease in $M_c$ by $\approx 0.005\, M_\odot$ due to EOS effects (Harpaz & Kovetz 1988) would *increase* $\Delta V$ ages by $\simeq 3\% - 4\%$, since it would make the HB appear fainter. The age reduction in Chaboyer & Kim 1995 (by $\simeq 6\% - 7\%$) is a *semi-empirical* effect, i.e., it is strictly valid only if a stellar evolution-independent calibration of the HB magnitudes is employed, or if the turnoff magnitude is estimated from some other means.

### 3.3. Reaction rates

The helium flash in the core of an RGB star occurs when the triple-$\alpha$ process, in which three $^4$He nuclei are transformed into a $^{12}$C nucleus, is ignited, with the building of a trace equilibrium abundance of $^8$Be nuclei (e.g., Clayton 1983; Castellani 1985; SG78).

Near $T \approx 10^8$ K, the rate of the triple-$\alpha$ reaction may be given as

$$\epsilon_{3\alpha} \propto T^{40} \tag{23}$$

[see, e.g., eq. (5-105) in Clayton 1983]. On the other hand, Sweigart (1994) has recently estimated that the maximum temperature within the degenerate helium core of an RGB star near the RGB tip increases at a rate $d\log T_{max}/dM_{sh} \simeq 1.4\, M_\odot^{-1}$, due to the increasing mass of hydrogen that is converted to helium at the hydrogen-burning shell. Thus, in order to delay the helium flash in such a way as to enable an increase in the core mass by, say, $0.015\, M_\odot$, an increase in the maximum core temperature by $\Delta \log T \approx 0.021$ should occur prior to helium ignition. This would imply that the present reaction rates are too *high* by $\Delta \log \epsilon_{3\alpha} \approx 40 \times 0.021 = 0.84$, i.e., a factor of $10^{0.84} \simeq 7$. [Sweigart's (1994) estimate of an error by a factor of 16 corresponds to an increase in the core mass by $\approx 0.02\, M_\odot$.] Thus, as pointed out by Sweigart (1994), quite a large error in the reaction rates would have to be present to cause an important increase in $M_c$.

The reaction rates most commonly employed for the two-step triple-$\alpha$ process are those given by Fowler, Caughlan, & Zimmerman (1975), Harris et al. (1983), Caughlan et al. (1985), and Caughlan & Fowler (1988). The latter two references give identical rates for this reaction, while some differences exist with respect to the others. It may be of some interest to reinvestigate the origin of such differences, since in the classical study by SG78 the earlier rates provided by Fowler et al. (1975) were used for the triple-$\alpha$ reactions.

We have computed the ratio between the more recent rates for the triple-$\alpha$ process and those provided by Fowler et al. (1975). For the conditions near the helium flash ($T \approx 10^8$ K), we find that

$$\epsilon_{3\alpha}(1988) \simeq 1.22 \times \epsilon_{3\alpha}(1975), \tag{24}$$

which implies a decrease in the core mass by only $\approx 0.002\, M_\odot$. The stated uncertainty in either reaction rate is somewhat larger than the difference between the two, albeit much smaller than, say, a factor of two. [To be sure, there is a second branch in the tabulated reaction rates which *is* uncertain by a large factor, but this becomes more important than the main branch only for $T \gtrsim 51 \times 10^8$ K, being thus irrelevant for the flash conditions.]

Oberhummer et al. (1994) have recently analyzed the triple-$\alpha$ process from a different perspective. These authors have shown that even extremely small variations in the strength of the effective nucleon-nucleon interaction (i.e., by $\approx 0.1\% - 0.2\%$) may have a large impact upon the reaction rates. The authors interpret their results in terms of the change that is introduced in the Boltzmann factors $\exp(-E_R/k_B T)$ due to the shifts in the resonance energies of the $^8$Be and $^{12}$C nuclei in the two steps of the reaction. For instance, a change in the resonance energy of the dominating $0_2^+$ state for the $^8$Be$(\alpha, \gamma)^{12}$C reaction from the present value of $E_R = 0.2875$ MeV to $E_R = 0.2517$ MeV would, by itself, change the reaction rate by a factor of 60 (cf. their Table 1). We estimate, from equation (4-195) of Clayton (1983), that a change by a factor of 7 in the reaction rate would require a smaller resonance energy for this state by $\approx 6\%$. Though this may seem small, we believe it is unlikely that the experimental determinations could be wrong by such an amount.

Finally, it is important to recall, after Tarbell & Rood (1975), that the so-called screening factors of the nuclear reactions are another relevant ingredient that influences the value of the core mass at the RGB tip, since they can considerably modify the unscreened nuclear reaction rates.

In modern stellar evolution computations, the screening factors usually employed are those by DeWitt, Graboske, & Cooper (1973) and Graboske et al. (1973), who have presented a general expression that is claimed to be valid in the so-called weak, intermediate, and strong screening regimes. More recently, Yakovlev & Shalybkov (1989) have readdressed in detail the problem of ion and electron screening in dense stellar interiors. These authors claim that their analysis is more accurate than the one by DeWitt et al. (1973) and Graboske et al. (1973), due to their use of more reliable calculations of the relevant Coulomb thermodynamic quantities. To our knowledge, the Yakovlev-Shalybkov prescriptions have not been incorporated into modern evolutionary computations, so that it would be worthwhile to compare these more recent results with those from Graboske et al. 1973.

For the present calculations, we assume, for simplicity, $\rho = 10^6 \, \mathrm{g \, cm^{-3}}$, $T = 10^8$ K, and $Y_c = 1$. For the gaseous, strongly electron-degenerate case, ion and electron screening for the triple-$\alpha$ process may be obtained from equations (184) and (186) in Yakovlev & Shalybkov 1989, respectively. For an enhancement factor given by

$$f = \exp(H), \tag{25}$$

we find, assuming *intermediate screening* to be operating,

$$H_i = 1.396, \quad f_i = 4.039, \tag{26}$$

where the subscript i denotes ion screening. As to electron screening, we find

$$H_e = 0.042, \quad f_e = 1.043. \tag{27}$$

Given that $f_e$ is very close to unity, we conclude that electron screening is weak.

Ion screening may also be obtained from equation (19) of Graboske et al. 1973, with the aid of their Table 4. Assuming $\theta_e = 0$ — which amounts to considering extreme degeneracy and no electron screening — the usual assumption in RGB interior calculations (cf. SG78), we find

$$H_i = 1.326, \quad f_i = 3.766. \tag{28}$$

[Note that the enhancement factor obtained from eq. (A6) in Tarbell & Rood 1975 for intermediate screening is $\simeq 2$ times larger than this result, even though their expression for weak screening, eq. (A5), agrees perfectly well with the value obtained from the Graboske et al. 1973 expression.]

This result is very similar to the one obtained from the Yakovlev & Shalybkov (1989) formalism, and we conclude that the differences between the Graboske et al. (1973) and Yakovlev & Shalybkov analyses are of minor relevance for the quantitative evaluation of $M_c$.

Unfortunately, the treatment of screening in the non-degenerate conditions that characterize the H-burning shell employed by Sweigart & Gross (SG78 and references therein) was not described in detail in their series of papers. Checking the influence of this ingredient in future calculations should also be of some interest.

### 3.4. Standard neutrino losses

Uncertainties in the standard fitting formulas for the neutrino emission rates are another potential source of error in the determination of $M_c$, as already emphasized by Chieffi & Straniero (1989). These authors note, for instance, that the standard energy loss rates due to photo, pair, and plasma processes, as given in the fitting formulas by Munakata, Kohyama, & Itoh (1985), are valid for the range $0 \leq \log \rho \leq 14$, $8 \leq \log T \leq 10$ whereas in RGB interiors, where typically $\log \rho \sim 6$, energy losses by neutrinos begin to be a relevant cooling mechanism well below $\log T = 8$. Note that evolutionary computations are commonly based upon the rates periodically improved by the Japanese group (cf. Itoh et al. 1992 and references therein).

Only very recently, though, a fitting formula has appeared (Haft, Raffelt, & Weiss 1994) that is sufficiently reliable for application in stellar evolution calculations aiming at a precise evaluation of $M_c$ (cf. Altherr & Salati 1994). Comparison of the new plasma emission rates with those previously obtained by several different groups reveals surprisingly large differences (cf. Figure 8 in Haft et al. 1994). Haft et al. have concluded that the *correct* emission rates in the parameter regime relevant for the RGB in GCs are larger by about $10\% - 20\%$ than those of (former) stellar evolution calculations, implying an increase in the core mass by $\simeq 0.005\ M_\odot$. The SG78 calculations, in particular, neglect neutral-current effects (Weinberg 1967; Salam 1968), which may have led to an underestimate of their $M_c$ values (Ramadurai 1976; SG78).

## 4. Non-standard evolutionary effects, standard physics: stellar rotation and microscopic diffusion

On the basis of the Mengel & Gross (1976) rotational models, Renzini (1977) has obtained the following relation between the core mass increase $\Delta M_c$ and the initial rate of stellar rotation $\omega$:

$$\Delta M_c = 1.25 \times 10^6 \, \omega^{2.16}, \qquad (29)$$

where $\omega$ is given in $\mathrm{rad\,s^{-1}}$. Thus, a stellar rotation rate as small as $\approx 2.2 \times 10^{-4}\,\mathrm{rad\,s^{-1}}$ might be sufficient to produce an increase in $M_c$ as large as $0.015\,M_\odot$.

The physical argument behind the Mengel & Gross (1976) results is quite simple: in the rotating stellar interior of an RGB star, besides degeneracy pressure working against further compression of the stellar material, there is also a centrifugal component working against the compression of the core and the eventual ignition of the triple-$\alpha$ reactions, thus retarding the helium flash and leading to an increase in $M_c$ over the non-rotating model predictions.

Following these theoretical arguments and based upon equation (29), much interpretative work on the properties of rotating RGB and (ensuing) HB structures has been developed (e.g., Renzini 1977; Fusi Pecci & Renzini 1976, 1978; Castellani, Ponte, & Tornambè 1980; Castellani & Tornambè 1981; Sweigart et al. 1987; Renzini & Fusi Pecci 1988; Shi 1995). Peterson (1983, 1985a,b) provided observational evidence that evolved GC stars do rotate while on the HB, and also suggested a very interesting trend of increasing rotation rate for clusters with bluer HB morphologies, which appeared to go in favor of a significant influence of rotation upon stellar evolution on the RGB and beyond.

Recently, more sophisticated stellar evolution codes that take into account the effects of rotation upon the evolving structure were developed. In these models, the influence of angular momentum loss from the stellar surface and of angular momentum redistribution within the stellar interior were consistently taken into account (Deliyannis, Demarque, & Pinsonneault 1989; Sofia, Pinsonneault, & Deliyannis 1990; Pinsonneault, Kawaler, & Demarque 1990; Pinsonneault, Deliyannis, & Demarque 1991a). This represented a clear improvement over the earlier models by Mengel & Gross (1976), where a simple one-dimensional approximation with angular momentum conservation in individual mass shells was adopted.

The predictions of these recent stellar evolution models were very different from those in the simple approximation by Mengel & Gross (1976). Deliyannis et al. (1989), in particular, in their Figs. 3 and 4, compare the amount of angular momentum remaining in their models near the zero-age main sequence and turnoff regions with the corresponding predictions in the Mengel & Gross case, concluding that the latter (but not the former) are inconsistent with the observed surface rotation velocities of HB stars and the observational limits for rotation near the turnoff. Most importantly for our present purposes, they point out that their models with differential rotation are unable to preserve sufficient angular momentum in the core as to cause a significant increase in $M_c$ over the values expected for non-rotating models, even though the core retains sufficient angular momentum that the observed HB velocities can be accounted for. They claim, as a consequence, that rotation cannot affect the core mass and the HB morphology in a significant way, and even speculate that the correlation obtained by Peterson (1985b) between stellar rotation of HB stars and HB morphology could be *driven by* metallicity and age effects.

It appears likely, however, that the issue has not been definitely resolved yet. The amount of initial angular momentum in population II stars, the angular momentum loss law and several physical instabilities that may be responsible for angular momentum redistribution in the stellar interior, are some of the aspects of the problem that may be improved upon. Besides, there are many free parameters in these models that must be adjusted in a semi-empirical way. Some of the most salient uncertainties have been pointed out by Michaud & Charbonneau (1991, especially their Sects. 6.6 and 7.5). We note as well that these models have not been evolved beyond the main-sequence turnoff region, so that the conclusions about more advanced phases have yet to be established in a more accurate way. For instance, the CNO abundances in RGB stars of GCs do not appear to be accounted for by these models without treating mixing processes that should take place on the RGB phase itself (Pinsonneault et al. 1991b). On the other hand, Sofia et al. (1990) have argued that since the rotational models do not appear to be in conflict with any particular observational data (including the light element abundances), any additional physical ingredient to be added should be of sufficiently small relevance. They also point out that many of the model results are not particularly sensitive to their choice of free parameters, this appearing to be the case also for the core mass at the flash (cf. Deliyannis et al. 1989).

With respect to the suggestion that age is a key variable responsible for the differences in rotational velocities among HB stars in GCs, there now appears to be at least one example that goes against this idea. While it is clear that M3 and M13 have very different stellar rotational velocities (Peterson 1985b) and HB morphologies for nearly identical metallicities, it has recently been argued that these two clusters do not have appreciably different ages (Catelan & de Freitas Pacheco 1995).[3]

It is well known that element diffusion may be responsible for generating sufficiently high molecular weight gradients as to inhibit rotational instabilities that provoke mixing (e.g., meridional circulation). In this case, the good agreement between rotational model predictions and the observations, in what concerns the light element abundances, might be rather fortuitous. In the case of low-mass stars, diffusion is mainly caused by gravitational settling, given that thermal diffusion is small for the conditions prevailing in such stars (cf. Deliyannis, Demarque, & Kawaler 1990), as is radiative levitation (Michaud, Fontaine, & Beaudet 1984).

Much work has recently been developed to investigate the impact of microscopic diffusion on stellar evolution predictions for population II stars. This has been motivated by the possibility of obtaining significant reductions in estimated GC turnoff ages (Stringfellow et al. 1983) and (in diffusion and rotation studies alike) by the need to estimate the cosmological lithium abundance from the observed Spite & Spite (1982) Li plateau for metal-poor dwarfs (Deliyannis et al. 1989, 1991).

---

[3]After this paper had been completed, the authors became acquainted with the latest investigation by Peterson, Rood, & Crocker (1995) on the correlation among stellar rotation, HB morphology, and abundances on the RGB. Their analysis of the clusters M13 and M3 appear to confirm our arguments.

However, apart from the discussion in Proffitt & VandenBerg (1991) regarding the impact of He diffusion upon the luminosities of HB stars and hence the ages that are estimated on the basis of the $\Delta V$ method, the influence of diffusion upon $M_c$ has not been studied in detail. Chaboyer et al. (1992), in their recent analysis of GC ages under helium diffusion, assume that the standard results remain valid in the diffusive approximation, provided a lower initial $Y$ is adopted. This assumption appears to be valid (see below) only insofar as the envelope helium abundance decreases uniformly as a function of [Fe/H] due to the influence of helium diffusion. Preliminary evidence in favor of this assumption has been presented by Proffitt & VandenBerg (1991), but more calculations for more metal-poor compositions are needed to corroborate their results.

Proffitt & VandenBerg (1991) argue that "the core mass that is ultimately attained at the helium flash is mainly a function of the envelope helium content," and also that "the evolution of a giant which has a decreased envelope helium abundance due to diffusion should be essentially the same as that which did not undergo gravitational settling but which simply had a lower $Y$ when it was formed." [Recall that the timescale for diffusion is such that the amount of He diffused is appreciable in relatively long-lived evolutionary phases only, but not on the RGB itself (cf. Deliyannis et al. 1990 for a detailed discussion).]

Assuming this argument to be correct, it follows that the change in $M_c$ due to microscopic diffusion can be obtained from the relation

$$\frac{\mathrm{d}M_c}{\mathrm{d}Y} = -0.24 \, M_\odot, \tag{30}$$

as prescribed by the SG78 and M89 models. From Fig. 3 in Proffitt & VandenBerg 1991, $Y_{\text{standard}} - Y_{\text{diffusion}} \in [0.006, 0.024]$, implying a $\Delta M_c / M_\odot \in [+0.00015, +0.006]$ for [Fe/H] = $-1.26$ and [O/Fe] = 0.55. Shi (1995) has recently claimed that "$M_c$ will be slightly increased by 0.005 $M_\odot$ with respect to that of standard models due to the helium settlement into the core." This is likely to be an upper limit, particularly in higher metallicity stars with thicker convection zones, since diffusion in convection zones is strongly inhibited.

Note that the Proffitt & VandenBerg (1991) argument is quite general, and may be extended to whatever processes that may lead to a change in the envelope helium abundance. For instance, meridional circulation, driven by rotational instabilities, may *increase* the envelope helium abundance of an RGB star and lead to a smaller $M_c$. Deliyannis et al. (1989) dismiss this (subtle) effect. Deep mixing, often invoked to explain abundance anomalies in RGB stars (see Freeman & Norris 1981 for a review), may play a similar rôle if started early enough on the RGB.

The interplay among these various mixing mechanisms is far from being well established, and it is possible indeed that mutual inhibitions take place. The effects of helium diffusion can inhibit rotational instabilities, which can inhibit diffusion, which in turn may be much less effective if overshooting or other type of convective instability is taking place (but see also Proffitt & Michaud 1991). Deliyannis & Demarque (1991) have concluded, on the basis of a comparison of their Li isochrones with the observed high-temperature end of the Li plateau, that their models *overestimate* diffusion in population II stars (but see also Proffitt & Michaud 1991 and Chaboyer et al. 1991), so that inhibiting mechanisms *must* be effective; Swenson (1995) has recently pointed out that mass loss on the main-sequence phase may improve the agreement, even though it is quite unclear whether the required rates, $\sim 10^{-12} M_\odot \, \mathrm{yr}^{-1}$, are attained in such stars. Much more theoretical and observational work is clearly necessary. Very recently, Chaboyer et al. (1995) have attempted to include rotation, diffusion, and overshooting simultaneously in their solar models, and found the final results to be significantly different from those achieved when these different processes are individually considered. When comparing their predictions of solar oscillation modes with the observations, they find indications of important missing ingredients in their treatment of rotation. Similar calculations should be carried out for population II stars.

One may find interesting that Figs. 1 and 2 suggest a simultaneous increase in $M_c$ and decrease in $Y$ — in the same sense as expected on the basis of equation (30) above. However, to account for an increase in $M_c$ by $0.015 \, M_\odot$, equation (30) would suggest a decrease in $Y$ by 0.06 — which is probably unrealistic.

To close the present section on the rôle of standard physical ingredients in non-standard evolutionary computations, we mention in passing that the recent results by Castellani & Castellani (1993) represent strong evidence against a significant influence of mass loss on the RGB upon $M_c$.

## 5. Non-standard evolutionary effects, non-standard physics

In this section, some representative novel physical ingredients are briefly discussed in the framework of RGB stellar evolution: a non-zero magnetic moment for the neutrino, the axion, and Weakly Interacting Massive Particles. Even though there are important reasons to regard these three exotic mechanisms as the main candidates for playing *actual* rôles in the evolution of stars, several other novel phenomena have been proposed whose impact on stellar evolution may be of relevance, such as millicharged particles (Davidson & Peskin 1994 and references therein), a "fifth force" (e.g., Grifols, Massó, & Peris 1989), etc., but which shall not be addressed here. The reader is referred to Raffelt 1990b for a critical discussion of some of these possibilities.

## 5.1. A non-zero neutrino magnetic moment

One proposed explanation for the "solar neutrino problem," motivated by a possible anticorrelation between solar magnetic activity and the measured solar neutrino flux on Earth (cf. Oakley et al. 1994 and references therein), is that electron neutrinos originating from the solar interior change their nature when traversing the solar convective zone. More specifically, it has been suggested that, due to an anomalously large magnetic moment, left-handed neutrinos would turn into right-handed neutrinos when reaching the magnetic field in the solar convective zone, and thus escape detection when reaching the Earth. Estimates by Voloshin, Vysotskii, & Okun (1986) suggest that a magnetic moment as high as $\mu_{\nu_e} \approx 10^{-10} \mu_B$ would be needed (where $\mu_B$ is Bohr's magneton), although more detailed considerations about the magnetic fields in the solar interior may allow smaller values for $\mu_{\nu_e}$ (Oakley et al. 1994).

Presently, laboratory experiments (cf. Vogel & Engel 1989 and references therein) appear unable to constrain the size of $\mu_{\nu_e}$ to better than an order of magnitude above this value. On the other hand, astrophysical arguments may be able to improve this bound. For instance, it has been claimed (e.g., Fukugita & Yazaki 1987) that the cosmological $^4$He abundance constrains the magnetic moment to $\mu_{\nu_e} < 5 \times 10^{-11} \mu_B$. Similarly, it has been argued that the right-handed neutrinos escaping SN1987A could be reflipped again to left-handed neutrinos, and then be detected on Earth: analysis of the neutrino signal from SN1987A, as monitored by the Kamiokande detectors, puts stringent constraints on the size of the neutrino magnetic moment (Nötzold 1988 and references therein), which however are model-dependent (see, for instance, Blinnikov & Okun 1988 and Vogel & Engel 1989 for discussions and additional references). Raffelt (1990b) provides an excellent review of the subject.

Another important astrophysical implication of a non-zero neutrino magnetic moment would be an increased rate of cooling of stellar degenerate cores. This is due to the increased (over the standard rates discussed in Sect. 3.4) energy losses from the dominant decay of plasmons $\gamma_{pl} \rightarrow \bar{\nu}_e \nu_e$ (Bernstein, Ruderman, & Feinberg 1963; Sutherland et al. 1976).

This aspect of the problem has been most recently addressed by Raffelt (1990a), Raffelt & Weiss (1992), and Castellani & Degl'Innocenti (1993). In particular, in the last two papers the effect of increased cooling due to the increased plasmon decay rate has been incorporated into detailed stellar evolution codes. Raffelt & Weiss (1992) compare their results with those by Castellani & Degl'Innocenti (1993), finding good agreement to within the approximations adopted in the latter study.

Both Raffelt & Weiss (1992) and Castellani & Degl'Innocenti (1993) point out that relations of the form depicted in Sect. 2 are in fact only first-order approximations to the real case. In fact, the relation between core mass and RGB luminosity actually depends sensitively on *the type of cooling* involved: the standard relation between these two quantities is not really universal. This implies that observational bounds on a possible core-mass excess will translate into slightly weaker bounds on $\mu_{\nu_e}$. Hence, we assume that the non-standard increase in $M_c$ from Sect. 2 translates into bounds on $\mu_{\nu_e}$ which are less restrictive by $\simeq 10^{-12}\mu_{\mathrm{B}}$ (Raffelt & Weiss 1992).

Employing Raffelt's (1990a) equation (45), corrected by Raffelt & Weiss (1992) to read [cf. their eq. (18)]

$$\Delta M_{\mathrm{c}} = 0.013 \, \frac{\mu_{\nu_e}}{10^{-12}} \mu_{\mathrm{B}}, \tag{31}$$

we find that a non-standard increase in $M_{\mathrm{c}}$ by $0.015\,M_{\odot}$ could be accounted for by a neutrino magnetic moment of

$$\mu_{\nu_e} \approx 1.15 \times 10^{-12} \, \mu_{\mathrm{B}}. \tag{32}$$

This value applies directly to the diagonal magnetic or electric dipole moments of Dirac neutrinos, or to the transition moments of Majorana neutrinos. For the transition moments of Dirac neutrinos, such a value should be multiplied by a factor $2^{-1/2}$ (Raffelt 1990b).

It is important to note that there appears to be no stronger semi-empirical constraint on the magnetic moment of the neutrino. In particular, Wang (1992a) and Blinnikov & Dunina-Barkovskaya (1992, 1993) have recently argued that the cooling of hot white dwarfs places the most reliable constraints on the neutrino magnetic moment, $\mu_{\nu_e} < 4 \times 10^{-12}\,\mu_{\mathrm{B}}$ (Wang 1992a), which would enable an increase in $M_{\mathrm{c}}$ by as much as $0.05\,M_{\odot}$. This is ruled out by Figs. 1 and 2 above. Nötzold's (1988) limit on $\mu_{\nu_e}$ from SN1987A considerations, $\mu_{\nu_e} < (1.5 \pm 0.5) \times 10^{-12}\,\mu_{\mathrm{B}}$, imply $\Delta M_{\mathrm{c}} < (0.020 \pm 0.007)\,M_{\odot}$.

Barbieri & Fiorentini (1987a,b) have recently argued, from simple theoretical considerations, that the neutrino magnetic moment should be smaller than $\mu_{\nu_e} \simeq 10^{-13}\,\mu_{\mathrm{B}}$. If this is true, neutrino magnetic moments produce negligible effects upon the He-core mass (i.e., $\Delta M_{\mathrm{c}} \lesssim 10^{-3}\,M_{\odot}$).

### 5.2. Axions

The axion is a hypothetical, (almost) massless particle, which appears as a pseudo-scalar Goldstone boson in a class of theoretical particle models. The axion gives an elegant solution to the problem of $CP$ conservation in strong interactions, and is a cold dark matter candidate (cf. Raffelt 1990b for a comprehensive review).

As discussed by Raffelt & Dearborn (1987), astrophysical considerations stand out as a unique opportunity of constraining the properties of the physical axion, since it may have a strong influence upon important evolutionary stages of stars, depending however on the details of the specific axion model.

*Hadronic axions* couple to ordinary nucleons but not to electrons, and arguments related to the solar age and solar helium abundance (Dearborn, Schramm, & Steigman 1986; Raffelt &

Dearborn 1987) constrain its mass to be

$$m_{\mathrm{a}} \lesssim 17 \,\mathrm{eV}\, R. \tag{33}$$

In a large class of grand unified theories, $R = 1$.

As in Sect. 5.1, we must check whether the standard core mass-luminosity relation is valid in the case of cooling of RGB interiors by axions. It is interesting to note that, contrary to the case in which only cooling by neutrinos is considered, cooling by axions produces a *central* helium flash (cf. Fig. 7e in Raffelt & Dearborn 1987).

From analysis of Table III in Raffelt & Dearborn 1987, we find that, for a 1.3 $M_\odot$ star, the axion mass which leads to an increase in $M_c$ by 0.069 $M_\odot$ ($m_{\mathrm{a}} \simeq 2.1$ eV) leads to an increase in the RGB tip luminosity by $\Delta \log L \simeq 0.34$. A similar increase in $M_c$, according to equation (10) in Raffelt 1990a, would lead to a brighter RGB tip by $\Delta \log L \simeq 0.324$. Since we are concerned with smaller core mass increments than the one considered in this specific example, we may neglect entirely this small difference: the *direct* effect of a possible axion cooling upon the RGB tip luminosity is therefore not relevant.

A very different effect, however, concerns the *duration of the helium-burning phase*. When we discussed neutrino cooling, we assumed that its sole implication upon helium-burning stars would be through its influence upon $M_c$. This is a sufficiently accurate approximation, since neutrino cooling is very inefficient for the physical conditions prevailing in the interior of an HB star. Energy losses by hadronic axions, on the other hand, dramatically *increase* while the star is on the HB (e.g., Raffelt 1990b), so that there is a further, *direct* influence upon the duration of the HB phase.

The following comparison illustrates the importance of this effect. The axionic evolutionary models by Raffelt & Dearborn (1987) show that an axion mass of $m_{\mathrm{a}} \simeq 2.1$ eV leads to an increase in the core mass by 0.069 $M_\odot$, and to a reduction in the duration of the HB phase by $\Delta \log t_{\mathrm{HB}} \simeq 0.88$. Equation (16) in Raffelt 1990a, on the other hand, indicates that the reduction due to the quoted increase in $M_c$ *alone* should be by only $\Delta \log t_{\mathrm{HB}} \simeq 0.20$. The difference is due to the fact that, in the latter relation, the *direct* impact of axions upon the HB phase has been neglected. This point has also been addressed by Haxton & Lee (1991).

We thus conclude that the relations obtained by Raffelt (1990a,b,c) involving the $R$-method arguments cannot be plainly employed to constrain a non-standard increase in $M_c$ (see also the discussion in Raffelt 1990b). However, the remaining relations employed in Sect. 2.1. can still be regarded as reasonable approximations to the core mass excess that might arise from axion emission on the RGB phase, which however should be checked against the HB lifetime constraint (see also Raffelt & Dearborn 1988).

Interpolating on Table III of Raffelt & Dearborn 1987, one may roughly estimate the required

axion mass to produce a $\Delta M_c = 0.015\,M_\odot$ to be

$$m_a \approx 0.6\,\text{eV}, \qquad \text{(hadronic axions)} \qquad (34)$$

which is more than one order of magnitude smaller than the upper bound inferred from the solar age argument (see above). From Table IV in Raffelt & Dearborn (1987), however, this is found to imply a reduction in the duration of the HB phase by $\approx 63\%$, which is rejected on the basis of star count arguments (e.g., Raffelt & Dearborn 1988). Therefore, hadronic axions cannot be entirely responsible for an increase in $M_c$ by as much as $\approx 0.015\,M_\odot$.

In addition to the hadronic axions, *DFSZ axions* (Dine, Fischler, & Srednicki 1981; Zhinitsky 1980) are believed to play an important rôle in particle physics. In the case of the DFSZ axions, which do couple to electrons, the emission occurs mainly through Compton and bremsstrahlung processes, while the Primakoff effect is the dominant emission process in the case of hadronic axions (Raffelt 1990b). The latest results by Raffelt & Weiss (1995) on the evolution of a low-mass, low-metallicity RGB star suggest that a core mass increase by $0.015\,M_\odot$ could be accounted for by DFSZ axions of mass

$$m_a \approx 7.4 \times 10^{-3}/\cos^2\beta\,\text{eV}. \qquad \text{(DFSZ axions)} \qquad (35)$$

Interestingly, the rate of increase in the pulsation period of the ZZ Ceti variable G117-B15A (Kepler et al. 1991), which appears too large in comparison with the expectations from CO white dwarf models, has motivated an accelerated cooling scenario (Isern, Hernanz, & Garcia-Berro 1992) in which an excess cooling by axions of an estimated mass $m_a \simeq (8 \pm 3)/\cos^2\beta$ meV might be taking place. However, recent results by Vauclair (1994) suggest that the variability is not as large as previously reported, and also that no obvious periodicity is present. Therefore, we believe that this argument offers no compelling evidence for axion emission, although this case is illustrative of the kind of observations that could lead to the positive discovery of non-standard cooling. It should be noted, in addition, that the axion emission rates employed by Isern et al. may have been underestimated (Altherr, Petitgirard, & del Río Gaztelurrutia 1994), so that their $m_a$ value may have been overestimated accordingly.

The mass value for the axion which would be needed to produce a $\Delta M_c \approx 0.015\,M_\odot$ is not excluded by analysis of the white dwarf luminosity function (Raffelt 1986; Raffelt & Weiss 1995; Altherr et al. 1994) nor by the limits imposed by type I supernova explosions (Wang 1992b). SN1987A considerations may not rule out $m_a$ values lying close to that found in equation (35) above either (Altherr et al. 1994).

### 5.3. Weakly Interacting Massive Particles

Another proposed candidate for the dark matter in the Universe are the so-called WIMPs. A wide variety of such non-baryonic, hypothetical particles has been suggested in the literature. A special class of WIMPs, termed "cosmions," has been proposed with properties which would help account for several astrophysical problems simultaneously: the solar neutrino problem; the age discrepancy between the oldest GCs and the measured expansion rate of our Universe; the dark matter problem; and, more recently, the "hump" in the GC stellar luminosity functions in the subgiant region (see, e.g., Faulkner 1991 for a review).

Since the existence of cosmions is presently only speculative, wide debate exists in the literature concerning their impact on stellar evolution. Different views on the problem have been published by Renzini (1987), Spergel & Faulkner (1988), Faulkner & Swenson (1988, 1993), Rood & Renzini (1990), Dearborn et al. (1990), Finzi (1991), Riazi (1991), Bouquet & Salati (1989), Bouquet, Kaplan, & Martin de Volnay (1989), Salati & Bouquet (1990), Martin de Volnay, Bouquet, & Kaplan (1990), Raffelt (1990d), etc. In general, these authors have computed the abundance of WIMPs in the stellar interior for different stellar evolutionary phases, assuming those particles to have been accreted from the dark Galactic component along the stellar lifetime.

We are mainly interested in the impact of efficient energy transport by WIMPs upon the low-mass red giants. Bouquet & Salati (1989) and Bouquet et al. (1989) have tackled in detail the problem of cosmion accretion and evaporation from stellar structures, and predicted that in the equilibrium condition cosmions will be so extremely concentrated in the RGB core, as a consequence of its much higher density in comparison with the more external layers, that no impact is expected upon RGB properties. Rood & Renzini (1990), on the other hand, have pointed out that a non-negligible influence upon the core evolution may be expected, given the non-local nature of WIMP energy transport. Since WIMPs are efficient energy carriers, a delay in the helium ignition may indeed be expected — one should recall that effects of order 1% in $M_c$ are of relevance for our present purposes.[4] Actual RGB evolutionary computations might be of interest in this regard. [Interestingly, Finzi's (1991) models suggest that a small core of hydrogen should develop at the very center of a WIMP-accreting RGB star, so that explosive pycnonuclear burning of hydrogen (and not thermonuclear burning of helium) would trigger the core flash, which would thus occur earlier and at a smaller $M_c$ value than in standard calculations.]

---

[4] Actually, the problem of cosmion evaporation from the stellar interior is complicated by the fact that the relevant WIMP cross sections are either purely axial or purely scalar (or vector) (cf. Faulkner & Swenson 1993). In the case that the cross-sections are purely axial (or spin-dependent), interaction with the stellar material would depend on the presence of protons, given that $\alpha$-particles are spinless and other nuclei with spin are much less abundant. In this case, transport effects by cosmions would decrease as hydrogen were converted to helium in the stellar interior, and thus little impact would be expected upon $M_c$ values. On the other hand, for scalar or vector interactions, transport effects would *increase* as the core grew richer in helium, so that a more significant impact on $M_c$ would appear reasonable.

The impact of WIMPs on the evolution of HB stars has been critically addressed by Renzini (1987), Spergel & Faulkner (1988), Dearborn et al. (1990), and Raffelt (1990d). In the latter two papers, evidence is found that the HB configurations become unstable in the presence of WIMPs, and that thermal pulses take place on short timescales until near core helium exhaustion. But little influence upon the HB lifetime has been found, in contrast with the simple arguments by Renzini (1987). Implications are also expected for RR Lyrae pulsation and the HB luminosity, as discussed by Dearborn et al. (1990) and Raffelt (1990d). These authors have found that the HB becomes brighter by $\simeq 0.3$ mag with respect to non-WIMPy models; however, the HB mass and metallicity explored by them are *not* very representative of GC stars, and evidence exists that the situation would be much more complex for stars that actually fall within the RR Lyrae instability strip. Note that the increase in HB brightness found by the above authors ensues *irrespective of an accompanying increase in* $M_c$, so that the relations obtained in Sect. 2 again cannot be straightforwardly employed when constraining the WIMP parameter space. At any rate, Spergel & Faulkner (1988) have noted that these calculations depend on assumptions about the evaporation of WIMPs during the helium flash, when the core expands in a very short timescale, and the central density drops from $\rho_c \approx 10^6 \, \mathrm{g \, cm^{-3}}$ (on the RGB tip) down to $\rho_c \approx 10^4 \, \mathrm{g \, cm^{-3}}$ (on the HB). Most of the authors who have considered the effect of WIMPs on the HB evolution have assumed that WIMPs do not evaporate in the flash.

Faulkner & Swenson (1993) have recently reviewed the evidence that the luminosity functions of well-observed GCs at the subgiant region cannot be reconciled with the predictions from standard stellar evolution. This topic has also been addressed quite recently by Stetson (1991), VandenBerg (1993), and VandenBerg & Stetson (1991). Interestingly, the theoretical luminosity functions of VandenBerg appear to actually favor the development of isothermal cores in GC stars as an explanation for the observed luminosity functions; Fig. 8 in Stetson 1991 is particularly impressive in this regard, showing that the "WIMPy" models lead to clearly improved matches with the observations in comparison with standard (as well as diffusive) evolutionary models. On the other hand, VandenBerg & Stetson (1991) have found that the "hook" feature in the CMD of the old open cluster M67 may not be accounted for by their WIMPy models, a discovery that has been termed by Faulkner (1991) as possibly "the most stringent stellar evolutionary test for the WIMP hypothesis." The hook feature is well accounted for by standard models.

In spite of the remarkable theoretical efforts that have been made by the above workers, it should be pointed out that the likelihood of cosmions as a solution to the solar neutrino problem (and thus to other problems in astrophysics) has decreased considerably in the recent years, as more data appeared on the solar neutrinos from the several independent groups now carrying out experiments in this field. Bahcall (1994) has reviewed the data for the $^7$Be and $^8$B neutrinos and argued very strongly against the "astrophysical solution" to the solar neutrino problem (i.e., a decreased central temperature leading to a smaller neutrino flux from the sun, as expected in the WIMP hypothesis). Schramm (1994) has also studied the problem in great detail, and even though his statistical analysis does not definitely rule out the astrophysical solution to the problem, it clearly favours explanations involving new neutrino physics — e.g., neutrino oscillations. Hata, Bludman, & Langacker (1994 and references therein) have strongly argued that the astrophysical solution is ruled out by comparison of the Kamiokande and Homestake solar neutrino data. The review by Berezinsky (1994) provides a useful entry to the recent literature, and in his Sect. 5 shows the trends expected in future, much more sensitive, solar neutrino experiments (SNO, Superkamiokande, Borexino) if the solar neutrino problem can be accounted for by the astrophysical solution — an idea against which Berezinsky as well strongly argues.

Several authors have also noted (e.g., Krauss 1990; Turner 1993) that the expected annihilation cross sections lead to a WIMP abundance that is simply too low (both within the sun and in the Universe as a whole) for the most "interesting" cosmion candidates, representing another extremely serious challenge to the cosmion hypothesis. In fact, recent developments do not even offer compelling evidence for the presence of significant amounts of non-baryonic dark matter in the scale of galactic halos, as reviewed by Turner (1993) and Schramm (1994, particularly his Fig. 5). In reality, this may be viewed as an argument against a large impact of axions upon stellar evolution in the Galaxy as well.

Irrespective of these difficulties to the WIMP hypothesis, one must recall that an independent satisfactory explanation to the anomalies in the luminosity functions of GCs has not been proposed as yet. Even if it turns out that WIMPs cannot be responsible for this anomaly, some other form of efficient central energy transport may have to be identified, if Faulkner & Swenson's (1993) arguments turn out to be correct. Implications for the core evolution of giants should then be studied in detail, checking whether an increase in $M_c$ over standard values is among the implied effects.

## 6. Conclusions

Revision and extension of Raffelt's (1990a,b,c) method and database to estimate a possible increase in helium-core mass values at the tip of the RGB over standard prescriptions from observational data and evolutionary models for RGB and HB stars suggest a $\Delta M_c \approx 0.01 \pm 0.015\ M_\odot$. Our $1 - \sigma$ limits on $\Delta M_c$ are higher than those previously obtained, and the ensuing constraints on novel physical mechanisms correspondingly weaker. Perhaps somewhat surprisingly, the method suggests that such an increase should be accompanied by a slight decrease in the envelope helium abundance. Similar results obtain for $\alpha$-enhanced compositions, assuming the Chieffi et al. (1991) scheme for treating the problem to be equally valid for RGB and HB stars. A high helium abundance for GCs, as recently advocated by Shi (1995), is ruled out by this method. A high $Y$ would also be in conflict with the explanation that Lee et al. (1990) have advanced for the Sandage period-shift effect (Sandage 1982, 1990, 1993), since RR Lyrae variables in Oosterhoff type II clusters would not be well-evolved objects, as required in this scenario. It would also produce pronounced blue loops in the evolutionary tracks (Sweigart & Gross 1976), which lead to "bifurcations" in synthetic HB models (e.g., Catelan 1992b, 1993a) that are not observed in real clusters. These important points have been missed in Shi's (1995) paper. The combination of a decreased $Y$ (Lee et al. 1990, Lee 1990) and an increased $M_c$ (Catelan 1992a,b, 1993a,b), on the other hand, would render the Yale evolutionary interpretation of the Sandage effect more attractive, although an explanation should then be found for this apparent anomaly in the helium abundance. A low primordial $Y$ seems to be ruled out, both on theoretical and observational grounds (e.g., Pagel et al. 1992; Pagel & Kazlauskas 1992; Olive et al. 1991; Olive & Steigman 1994).

Several aspects of the input physics involved in evolutionary calculations which may bear a direct influence upon $M_c$ values have been reviewed in the present article, with the purpose of identifying likely causes for a possible core mass increase. In particular, the following aspects have been discussed: conduction properties of dense stellar matter; Coulomb effects upon the EOS; reaction rates; cooling by neutrinos and the influence of its (possible) non-zero magnetic moment; stellar rotation and diffusion; energy losses by hadronic and DFSZ axions and by WIMPs.

Our analysis reveals that some of these aspects have already been sufficiently elaborated in recent work that an increase in $M_c$ by $\approx 0.005\ M_\odot$ is already indicated, as in the case of neutrino cooling (Haft et al. 1994). Efficient cooling by exotic particles (or some other mechanism) should also be considered seriously, after the challenging results of Faulkner & Swenson (1993) on the luminosity function of well-studied GCs: in particular, standard models appear incapable of accounting for the observed number of stars in the sub-giant region of the color-magnitude diagram. Uncertainties in the reaction rates appear small enough that changes in $M_c$ due to them may safely be regarded small.[5] Screening in the interior of an RGB star is also a secondary source of uncertainty; however, we were unable to ascertain the influence of screening upon the hydrogen-burning reactions taking place in the vicinity of the helium core. Although recent results tend to place rotation as an irrelevant ingredient insofar as $M_c$ values are concerned (Deliyannis et al. 1989), several sources of uncertainty still exist which would recommend caution before regarding this the final word on the subject. In particular, it is to be noted that the "age-driven" rotation effect on HB stars that has been speculated by these authors as an explanation to Peterson's (1983, 1985a,b) observed correlation between stellar rotation rate and HB morphology must be discarded if M13 (blue HB morphology and fast HB rotators) and M3 (even HB morphology and slower HB rotators) are really essentially coeval, as recently claimed by Catelan & de Freitas Pacheco (1995). Any effect connected to rotation or diffusion and leading to a change in surface helium abundances is also expected to lead to a change in the core mass value at the RGB tip; there is still much uncertainty in this field, where useful work could still be carried out. We point out, in addition, that one of the most commonly employed sources of conductive opacities for stellar interior calculations, I$^3$M83, actually refers to the liquid metal phase ($\Gamma \geq 2$) only, whereas in the RGB core one hardly achieves $\Gamma > 0.8$. Thus, an uncertain extrapolation from a particular physical phase (liquid) to a different one (gas) is usually carried out. It is also worth noting that I$^3$M83 have neglected the $e - e$ contribution to the conductive opacities (Blinnikov & Dunina-Barkovskaya 1992), an inadequate approximation for the RGB interior. To the present authors, the results of HL69 would appear more reliable *for the calculation of population II RGB sequences*, but care should be taken to employ sufficiently reliable interpolation schemes: we find the Ib75 approximation formulas to underestimate the HL69 opacities in the regions of interest for RGB computations. It would be of great interest to extend the results by the Japanese group toward smaller $\Gamma$. Employing the I$^3$M83 expressions *in their present form* to RGB interiors, a larger core mass is expected in comparison with those based upon HL69, which may help explain the disagreement between the SG78 and M89 results. Finally, we call attention to the fact that the Debye-Hückel correction to the EOS is not valid for the dense conditions characterizing RGB cores, so that a more refined treatment is required for the computation of accurate $M_c$ values.

_______________

[5]The surprisingly different reaction rates recently reported by Fowler (1993) for the triple-$\alpha$ process are probably due to his consideration of a very different physical environment.

The authors are indebted to Drs. Georg Raffelt, Allen Sweigart, Ben Dorman, Don VandenBerg, and an anonymous referee for carefully reading drafts of this paper and the manuscript and for providing many important comments and/or suggestions. M. C. acknowledges financial support by FAPESP (grant 92/2747-8). J. E. H. has been partially supported by CNPq.

# REFERENCES

Alongi, M., Bertelli, G., Bressan, A., & Chiosi, C. 1991, A&A, 244, 95

Altherr, T., Petitgirard, E., & del Río Gaztelurrutia, T. 1994, Astropart. Phys., 2, 157

Altherr, T., & Salati, P. 1994, Nucl. Phys. B, 421, 662

Bahcall, J. N. 1994, in: Salamfestschrift, ed. A. Ali, J. Ellis, & S. Randjbar-Daemi, World Scientific, Singapore, p. 213

Bahcall, J. N., & Pinsonneault, M. H. 1995, Rev. Mod. Phys., in press

Barbieri, R., & Fiorentini, G. 1987a, in: Neutrino Masses and Neutrino Astrophysics, ed. V. Barger, F. Halzen, M. Marshak, & K. Olive, World Scientific, Singapore, p. 140

Barbieri, R., & Fiorentini, G. 1987b, Nucl. Phys. B, 304, 909

Berezinsky, V. 1994, Comm. Nucl. Part. Phys., 21, 249

Bernstein, J., Ruderman, M., & Feinberg, G. 1963, Phys. Rev., 132, 1227

Blinnikov, S. I., & Dunina-Barkovskaya, N. V. 1992, MNRAS, 266, 289

Blinnikov, S. I., & Dunina-Barkovskaya, N. V. 1993, Astron. Rep., 37, 187

Blinnikov, S. I., & Okun, L. B. 1988, Sov. Astr. Lett., 14, 368

Bono, G., & Stellingwerf, R. F. 1994, ApJS, 93, 233

Bouquet, A., Kaplan, J., & Martin de Volnay, F. 1989, A&A, 222, 103

Bouquet, A., & Salati, P. 1989, A&A, 217, 270

Buonanno, R., Corsi, C. E., & Fusi Pecci, F. 1989, A&A, 216, 80

Buzzoni, A., Fusi Pecci, F., Buonanno, R., & Corsi, C. E. 1983, A&A, 128, 94

Cacciari, C., & Bruzzi, A. 1993, A&A, 276, 87

Canuto, V. 1970, ApJ, 159, 641

Caputo, F., De Rinaldis, A., Manteiga, M., Pulone, L., & Quarta, M. L. 1993, A&A, 276, 41

Caputo, F., & De Santis, R. 1992, AJ, 104, 253

Caputo, F., Martinez Roger, C., & Paez, E. 1987, A&A, 183, 228

Carney, B. W., Storm, J., & Jones, R. V. 1992, ApJ, 386, 663

Castellani, V. 1985, Fund. Cosm. Phys., 9, 317

Castellani, M., & Castellani, V. 1993, ApJ, 407, 649

Castellani, V., Chieffi, A., & Pulone, L. 1991, ApJS, 76, 911

Castellani, V., & De Santis, R. 1994, ApJ, 430, 624

Castellani, V., & Degl'Innocenti, S. 1993, ApJ, 402, 574

Castellani, V., Ponte, G., & Tornambè, A. 1980, Ap&SS, 73, 11

Castellani, V., & Tornambè, A. 1981, A&A, 96, 207

Catelan, M. 1992a, A&A, 261, 443

Catelan, M. 1992b, A&A, 261, 457

Catelan, M. 1993a, A&AS, 98, 547

Catelan, M. 1993b, in: New Perspectives on Stellar Pulsation and Pulsating Variable Stars, ed. J. M. Nemec & J. M. Matthews, Proc. IAU Colloq. 139, Cambridge University Press, Cambridge, p. 280

Catelan, M. 1995, in preparation

Catelan, M., & de Freitas Pacheco, J. A. 1995, A&A, 297, 345

Caughlan, G. R., & Fowler, W. A. 1988, Atomic Data Nucl. Data, 40, 283

Caughlan, G. R., Fowler, W. A., Harris, M. J., & Zimmerman, B. A. 1985, Atomic Data Nucl. Data, 32, 197

Chaboyer, B., Deliyannis, C. P., Demarque, P., Pinsonneault, M. H., & Sarajedini, A. 1991, poster paper presented at the 177[th] meeting of the American Astronomical Society; abstract in BAAS, 22, 1205

Chaboyer, B., Demarque, P., Guenther, D. B., & Pinsonneault, M. H. 1995, ApJ, in press

Chaboyer, B., & Kim, Y.-C. 1995, preprint

Chaboyer, B., Sarajedini, A., & Demarque, P. 1992, ApJ, 394, 515

Chieffi, A., & Straniero, O. 1989, ApJS, 71, 47

Chieffi, A., Straniero, O., & Salaris, M. 1991, in: The Formation and Evolution of Star Clusters, ed. K. Janes, ASP Conference Series, 13, p. 219

Clayton, D. D. 1983, Principles of Stellar Evolution and Nucleosynthesis. University of Chicago, Chicago

Crotts, A. P. S., Kunkel, W. E., & Heathcote, S. R. 1995, ApJ, 438, 724

Davidson, S., & Peskin, M. 1994, Phys. Rev. D, 49, 2114

Dearborn, D. S. P., Raffelt, G., Salati, P., Silk, J., and Bouquet, A. 1990, Nat, 343, 347

Dearborn, D. S. P., Schramm, D. N., & Steigman, G. 1986, Phys. Rev. Lett., 56, 26

Debye, P., & Hückel, E. 1923, Physik. Z., 24, 185

Deliyannis, C. P., & Demarque, P. 1991, ApJ, 379, 216

Deliyannis, C. P., Demarque, P., & Kawaler, S. D. 1990, ApJS, 73, 21

Deliyannis, C. P., Demarque, P., Kawaler, S. D., Krauss, L. M., & Romanelli, P. 1989, Phys. Rev. Lett., 62, 1583

Deliyannis, C. P., Demarque, P., & Pinsonneault, M. H. 1989, ApJ, 347, L73

Deliyannis, C. P., Demarque, P., & Pinsonneault, M. H. 1991, BAAS, 22, 1214

Deupree, R. G., & Wallace, R. K. 1987, ApJ, 317, 724

DeWitt, H. E., Graboske, H. C., & Cooper, M. S. 1973, ApJ, 181, 439

Dine, M., Fischler, W., & Srednicki, M. 1981, Phys. Lett. B, 104, 199

Dorman, B. 1992, ApJS, 81, 221

Faulkner, J. 1991, in: Challenges to Theories of the Structure of Moderate-Mass Stars, ed. D. Gough & J. Toomre, Lecture Notes in Physics, 388, p. 293

Faulkner, J., & Swenson, F. J. 1988, ApJ, 329, L47

Faulkner, J., & Swenson, F. J. 1993, ApJ, 411, 200

Fernley, J. 1994, A&A, 284, L16

Finzi, A. 1991, A&A, 247, 261

Fowler, W. A. 1993, Phys. Rep., 227, 313

Fowler, W. A., Caughlan, G. R., & Zimmerman, B. A. 1975, ARA&A, 13, 69

Freeman, K. C., & Norris, J. 1981, ARA&A, 19, 319

Frogel, J. A., Persson, S. E., & Cohen, J. G. 1983, ApJS, 53, 713

Fukugita, M., & Yazaki, S. 1987, Phys. Rev. D, 37, 3817

Fusi Pecci, F., & Renzini, A. 1976, A&A, 46, 447

Fusi Pecci, F., & Renzini, A. 1978, in: The HR Diagram, ed. A. G. D. Philip & D. S. Hayes, Reidel, Dordrecht, p. 225

Graboske, H. C., DeWitt, H. E., Grossman, A. S., & Cooper, M. S. 1973, ApJ, 181, 457

Grifols, J. A., Massó, E., & Peris, S. 1989, Mod. Phys. Lett. A, 4, 311

Guenther, D. B., Demarque, P., Kim, Y.-C., & Pinsonneault, M. H. 1992, ApJ, 387, 372

Haft, M., Raffelt, G., & Weiss, A. 1994, ApJ, 425, 222; erratum: 1995, ApJ, 438, 1017

Härm, R., & Schwarzschild, M. 1966, ApJ, 145, 496

Harpaz, A., & Kovetz, A. 1988, ApJ, 331, 898

Harris, M. J., Fowler, W. A., Caughlan, G. R., & Zimmerman, B. A. 1983, ARA&A, 21, 165

Hata, N., Bludman, S., & Langacker, P. 1994, Phys. Rev. D, 49, 3622

Haxton, W. C., & Lee, K. Y. 1991, Phys. Rev. Lett., 66, 2557

Hubbard, W. B., & Lampe, M. 1969, ApJS, 163, 297 (HL69)

Iben, I., Jr. 1968a, ApJ, 154, 557

Iben, I., Jr. 1968b, Nature, 220, 143

Iben, I., Jr. 1975, ApJ, 196, 525 (Ib75)

Iben, I., Jr., & Renzini, A. 1984, Phys. Rep., 105, 329

Isern, J., Hernanz, M., & Garcia-Berro, E. 1992, ApJ, 392, L23

Itoh, N. 1992, Rev. Mex. Astron. Astrof., 23, 231

Itoh, N., Hayashi, H., & Kohyama, Y. 1993, ApJ, 418, 405

Itoh, N., & Kohyama, Y. 1993, ApJ, 404, 268

Itoh, N., Kohyama, Y., Matsumoto, N., & Seki, M. 1984, ApJ, 285, 758; erratum: 1993, ApJ, 404, 418

Itoh, N., Mitake, S., Iyetomi, H., & Ichimaru, S. 1983, ApJ, 273, 774 ($I^3M83$)

Itoh, N., Mutoh, H., Hikita, A., & Kohyama, Y. 1992, ApJ, 395, 622

Kepler, S. O., et al. 1991, ApJ, 378, L45

Krauss, L. M. 1990, in: Particle Astrophysics: The Early Universe and Cosmic Structures, ed. J.-M. Alimi, A. Blanchard, A. Bouquet, F. Martin de Volnay, & J. Tran Thanh Van, Editions Frontières, Gif-sur-Yvette, p. 341

Lang, K. R. 1986, Astrophysical Formulae. Springer-Verlag, Berlin

Lee, M. G., Freedman, W. L., & Madore, B. F. 1993, ApJ, 417, 553

Lee, Y.-W. 1990, ApJ, 363, 159

Lee, Y.-W, Demarque, P., & Zinn, R. 1990, ApJ, 350, 155

Lee, Y.-W, Demarque, P., & Zinn, R. 1994, ApJ, 423, 248

Martin de Volnay, F., Bouquet, A., & Kaplan, J. 1990, in: Particle Astrophysics: The Early Universe and Cosmic Structures, ed. J.-M. Alimi, A. Blanchard, A. Bouquet, F. Martin de Volnay, & J. Tran Thanh Van, Editions Frontières, Gif-sur-Yvette, p. 265

Mathews, G. J., Schramm, D. N., & Meyer, B. S. 1993, ApJ, 404, 476

Mazzitelli, I. 1989, ApJ, 340, 249 (M89)

Mengel, J. G., & Gross, P. G. 1976, Ap&SS, 41, 407

Michaud, G., & Charbonneau, P. 1991, Sp. Sc. Rev., 57, 1

Michaud, G., Fontaine, G., & Beaudet, G. 1984, ApJ, 282, 206

Mitake, S., Ichimaru, S., & Itoh, N. 1984, ApJ, 277, 375

Munakata, H., Kohyama, Y., & Itoh, N. 1985, ApJ, 296, 197

Nandkumar, R., & Pethick, C. J. 1984, MNRAS, 209, 511

Nötzold, D. 1988, Phys. Rev. D, 38, 1658

Oakley, D. S., Snodgrass, H. B., Ulrich, R. K., & VanDeKop, T. L., 1994, ApJ, 437, L63

Oberhummer, H., Krauss, H., Grün, K., Rauscher, T., Abele, H., Mohr, P., & Staudt, G. 1994, preprint

Olive, K. A., & Steigman, G. 1994, preprint (UMN-TH-1230/94)

Olive, K. A., Steigman, G., & Walker, T. P., 1991, ApJ, 380, L1

Pagel, B. E. J., & Kazlauskas, A. 1992, MNRAS, 256, 49p

Pagel, B. E. J., Simonson, E. A., Terlevich, R. J., Edmunds, M. G. 1992, MNRAS, 255, 325

Panagia, N., Gilmozzi, R., Macchetto, F., Adorf, H.-M., & Kirshner, R. P. 1991, ApJ, 380, L23; erratum: 1992, ApJ, 386, L31

Peterson, R. C. 1983, ApJ, 275, 737

Peterson, R. C. 1985a, ApJ, 289, 320

Peterson, R. C. 1985b, ApJ, 294, L35

Peterson, R. C., Rood, R. T., & Crocker, D. A. 1995, ApJ, in press

Pinsonneault, M. H., Deliyannis, C. P., & Demarque, P. 1991a, ApJ, 367, 239

Pinsonneault, M. H., Demarque, P., Sofia, S., & Deliyannis, C. P. 1991b, BAAS, 22, 1206

Pinsonneault, M. H., Kawaler S. D., & Demarque, P. 1990, ApJS, 74, 501

Proffitt, C. R., & Michaud, G. 1991, ApJ, 371, 584

Proffitt, C. R., & VandenBerg, D. A. 1991, ApJS, 77, 473

Raffelt, G. G. 1986, Phys. Lett. B, 166, 402

Raffelt, G. G. 1990a, ApJ, 365, 559

Raffelt, G. G. 1990b, Phys. Rep., 198, 1

Raffelt, G. G. 1990c, Phys. Rev. Lett., 64, 2856

Raffelt, G. G. 1990d, in: Particle Astrophysics: The Early Universe and Cosmic Structures, ed. J.-M. Alimi, A. Blanchard, A. Bouquet, F. Martin de Volnay, & J. Tran Thanh Van, Editions Frontières, Gif-sur-Yvette, p. 283

Raffelt, G. G., & Dearborn, D. S. P. 1987, Phys. Rev. D, 36, 2211

Raffelt, G. G., & Dearborn, D. S. P. 1988, Phys. Rev. D, 37, 549

Raffelt, G. G., & Weiss, A. 1992, A&A, 264, 536

Raffelt, G. G., & Weiss, A. 1995, Phys. Rev. D, 51, 1495

Ramadurai, S. 1976, MNRAS, 176, 9

Renzini, A. 1977, in: Advanced Stages of Stellar Evolution, ed. P. Bouvier, A. Maeder, Geneva Obs., Geneva, p. 149

Renzini, A. 1987, A&A, 171, 121

Renzini, A., & Fusi Pecci, F. 1988, ARA&A, 26, 199

Riazi, N. 1991, MNRAS, 248, 555

Richer, H. B., & Fahlman, G. C. 1986, ApJ, 304, 273

Rogers, F. J. 1994, in: The Equation of State in Astrophysics, ed. G. Chabrier & E. Schatzman, Proc. IAU Colloq. 147, Cambridge University Press, Cambridge, p. 16

Rood, R. T., & Renzini, A. 1990, in: Astronomy, Cosmology and Fundamental Physics, ed. M. Caffo et al., Kluwer, Dordrecht, p. 287

Sackman, I.-J., Boothroyd, A. I., & Kraemer, K. E. 1993, ApJ, 418, 457

Saha, A., Freedman, W. L., Hoessel, J. G., & Mossman, A. E. 1992, AJ, 104, 1072

Salam, A. 1968, Elementary Particle Physics, ed. N. Svartholm. Almquist & Wikese, Stockholm

Salaris, M., Chieffi, A., & Straniero, O. 1993, ApJ, 414, 580

Salati, P., & Bouquet, A. 1990, in: Particle Astrophysics: The Early Universe and Cosmic Structures, ed. J.-M. Alimi, A. Blanchard, A. Bouquet, F. Martin de Volnay, & J. Tran Thanh Van, Editions Frontières, Gif-sur-Yvette, p. 273

Sandage, A. 1982, ApJ, 252, 553

Sandage, A. 1990, ApJ, 350, 603

Sandage, A. 1993, AJ, 106, 719

Schramm, D. N. 1994, in: Salamfestschrift, ed. A. Ali, J. Ellis, & S. Randjbar-Daemi, World Scientific, Singapore, p. 221

Shi, X. 1995, ApJ, 446, 637

Simon, N. R., & Clement, C. M. 1993, ApJ, 410, 526

Sofia, S., Pinsonneault M. H., & Deliyannis, C. P. 1990, in: Angular Momentum Evolution of Young Stars, NATO Advanced Research Workshop

Spergel, D. N., & Faulkner, J. 1988, ApJ, 331, L21

Spite, F., & Spite, M. 1982, A&A, 115, 357

Stetson, P. B. 1991, in: The Formation and Evolution of Star Clusters, ed. K. Janes, ASP Conference Series, 13, p. 88

Storm, J., Carney, B. W., & Jones, R. V. 1994, A&A, 290, 443

Straniero, O. 1988, A&AS, 76, 157

Straniero, O., & Chieffi, A. 1991, ApJS, 76, 525

Stringfellow, G. S., Bodenheimer, P., Noerdingler, P. D., & Arigo, R. J. 1983, ApJ, 264, 228

Sutherland, P., Ng, J. N., Flowers, E., Ruderman, M., & Inman, C. 1976, Phys. Rev. D, 13, 2700

Sweigart, A. V. 1994, ApJ, 426, 612

Sweigart, A. V., & Gross, P. G. 1976, ApJS, 32, 367

Sweigart, A. V., & Gross, P. G. 1978, ApJS, 36, 405 (SG78)

Sweigart, A. V., Renzini, A., & Tornambè, A. 1987, ApJ, 312, 762

Swenson, F. J. 1995, ApJ, 438, L87

Tarbell, T. D., & Rood, R. T. 1975, ApJ, 199, 443

Timmes, F. X. 1992, ApJ, 390, L107

Turner, M. S. 1993, Proc. Nat. Acad. Sci. U.S.A., 90, 4827

Urpin, V. A., & Yakovlev, D. G. 1980, Soviet Ast., 24, 126

van den Bergh, S. 1995, ApJ, 446, 39

VandenBerg, D. A. 1992a, ApJ, 391, 685

VandenBerg, D. A. 1992b, private communication

VandenBerg, D. A. 1993, in: The Globular Cluster - Galaxy Connection, ed. G. H. Smith & J. P. Brodie, ASP Conference Series, 48, p. 230

VandenBerg, D. A., & Stetson, P. B. 1991, in: Challenges to Theories of the Structure of Moderate-Mass Stars, ed. D. Gough & J. Toomre, Lecture Notes in Physics, 388, p. 367

Vauclair, G. 1994, private communication

Vogel, P., & Engel, J. 1989, Phys. Rev. D, 39, 3378

Voloshin, M. B., Vysotskii, M. I., & Okun, L. B. 1986, Sov. Phys. JETP, 64, 446

Walker, A. R. 1992a, ApJ, 390, L81

Walker, A. R. 1992b, AJ, 104, 1395

Wang, J. 1992a, Ap&SS, 189, 1

Wang, J. 1992b, Phys. Lett. B, 291, 97

Weinberg, S. 1967, Phys. Rev. Lett., 19, 1264

Wheeler, J. C., Sneden, C., & Truran, J. W., Jr. 1989, ARA&A, 27, 279

Yakovlev, D. G., & Shalybkov, D. A. 1989, Sov. Sci. Rev. E, 7, 311

Yakovlev, D. G., & Urpin, V. A. 1980, Soviet Ast., 24, 303

Zhitnitsky, A. R. 1980, Sov. J. Nucl. Phys., 31, 260

Fig. 1.— Graphical method to estimate $Y$ and $\Delta M_c$ from the models and the observations. Four bands are displayed, obtained from analysis of the quantities $M_V^{\rm RR}$, $\Delta M_V^{\rm tip-RR}$, $R$, and $A_{3.85}$ as described in the text. There is a relatively narrow "compatibility region" in the middle of the diagram, around $(Y, \Delta M_c/M_\odot) \approx (0.221, +0.011)$, where all of the four bands intersect, which thus defines the preferred solutions for $Y$ and $\Delta M_c$.

Fig. 2.— As in Fig. 1, except that an overall enhancement of the $\alpha$-elements, $[\alpha/{\rm Fe}] = 0.48$, has been assumed. Note that the "compatibility region" is around $(Y, \Delta M_c/M_\odot) \approx (0.223, +0.008)$.

Fig. 3.— Comparison among several different prescriptions for the conductive opacities along the evolution of the core properties of an RGB star. In panel (a), the conductive opacities are given, having been obtained from the indicated studies. In panel (b), the opacities are compared with the reference values from the Ib75 approximation to the HL69 results. The curves are plotted for a star with evolutionary parameters $(M/M_\odot, Y_{\rm MS}, Z) = (0.9, 0.20, 10^{-4})$. The core densities (given in cgs units) and temperatures were obtained from SG78. We have assumed a pure-helium plasma. The thick solid lines display the evolution of $\Gamma$, a non-dimensional parameter that measures the strength of electrostatic interactions in the medium ($\Gamma \lesssim 2$ for a gas, and $\Gamma \gtrsim 178$ for a crystal); and $\Lambda$, which measures the range of validity ($\Lambda \lesssim 0.2$) of the Debye-Hückel approximation to the Coulomb effects upon the EOS.

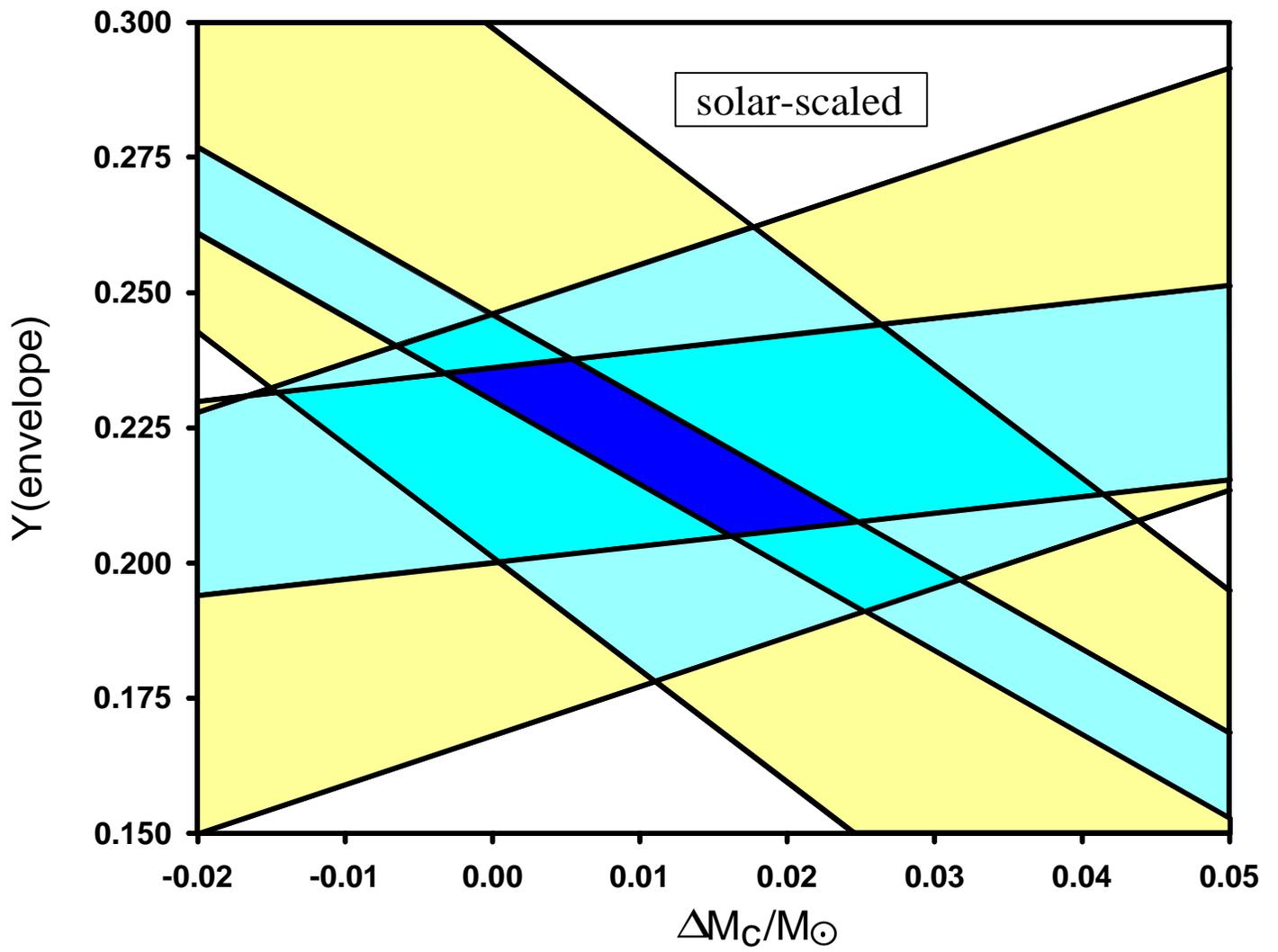

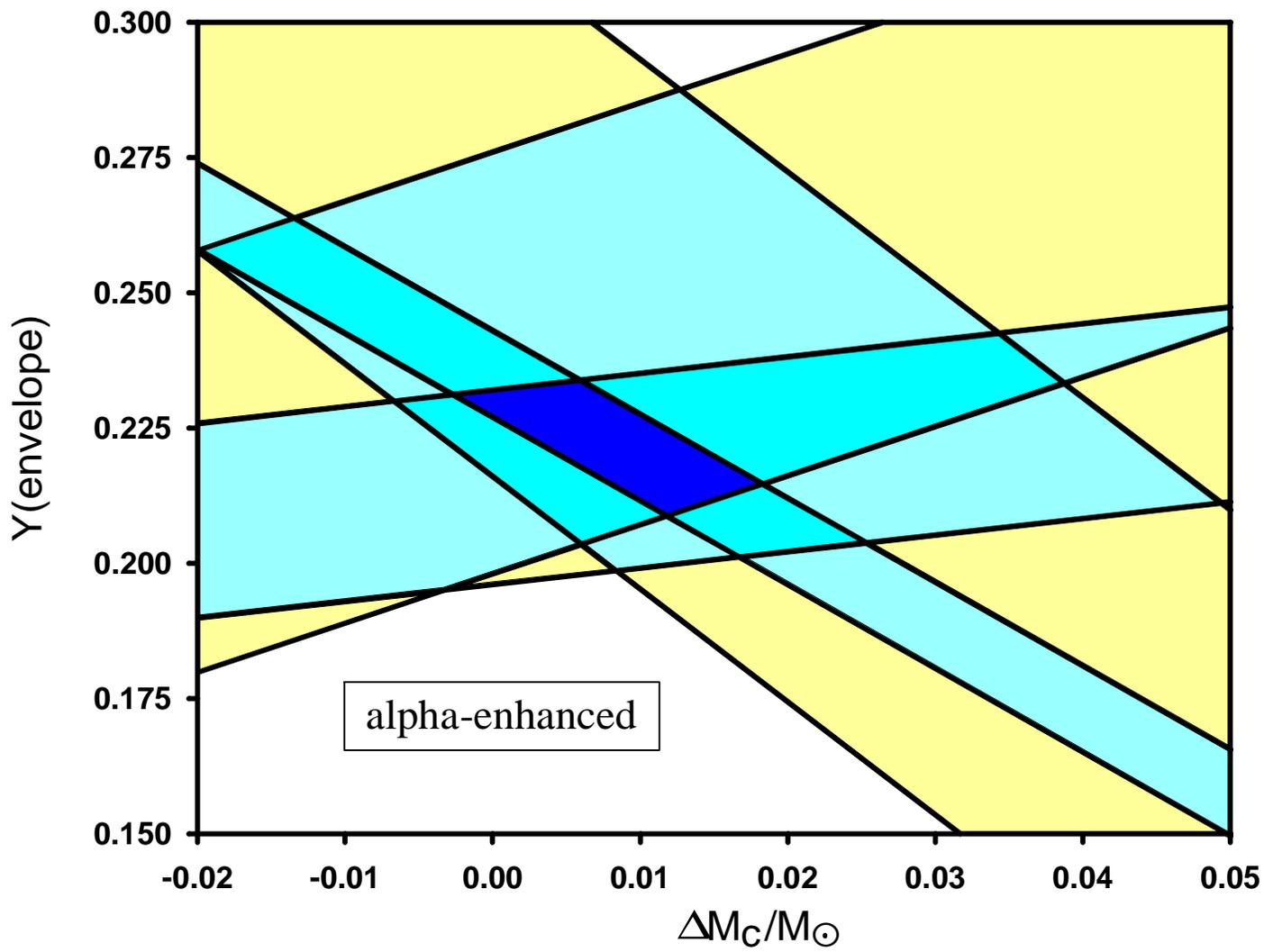

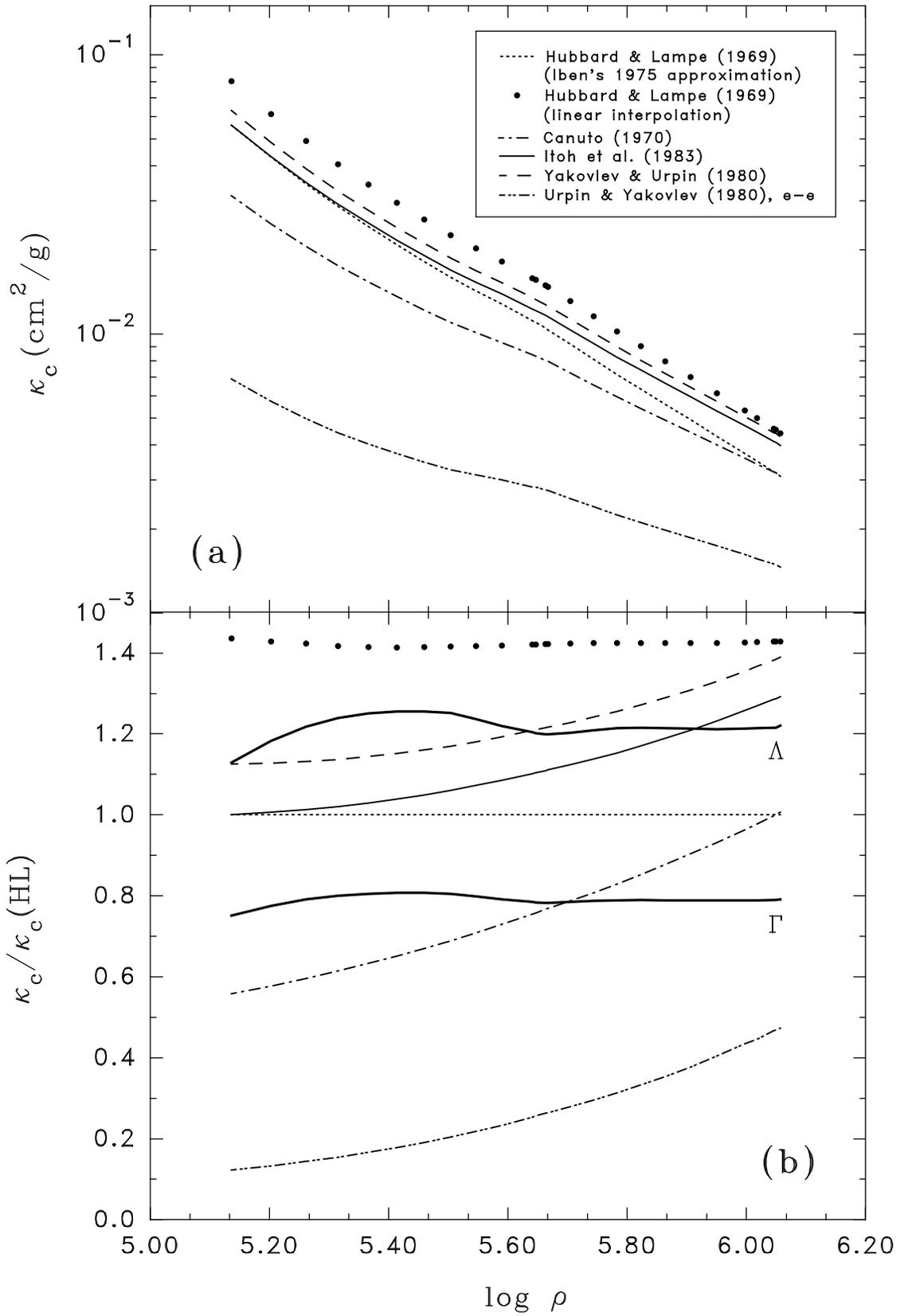